\documentclass[a4paper,11pt]{article}
\usepackage{jcappub} 
\usepackage{lineno}
\usepackage{orcidlink} 
\usepackage{xspace}
\usepackage[dvipsnames]{xcolor}
\usepackage{float}
\usepackage{subcaption}
\usepackage[percent]{overpic}
\usepackage[table,dvipsnames]{xcolor}
\usepackage{tabularx}
\usepackage[normalem]{ulem} 
\usepackage{multirow}
\usepackage{booktabs}
\usepackage[table]{xcolor}
\usepackage{amsmath} 
\usepackage{array}
\usepackage{colortbl}
\usetikzlibrary{shapes,arrows,positioning,calc}
\usepackage{xcolor}

\usepackage{tikz}  
\usepackage{tikz-3dplot} 
\usetikzlibrary{calc,3d}

\usetikzlibrary{shapes.geometric, arrows, positioning, fit}
\definecolor{lightgray}{gray}{0.93}
\definecolor{headergray}{gray}{0.82}
\usetikzlibrary{shadows, arrows.meta}

\newcolumntype{M}[1]{>{\centering\arraybackslash}m{#1}}

\usepackage{tikz}
\usetikzlibrary{
  matrix,          
  calc,            
  arrows.meta,     
  positioning,     
  shapes.geometric
}

\definecolor{crisp}{RGB}{200, 200, 200}
\colorlet{Mycolor1}{green!10!orange}
\usepackage{tcolorbox}
\tcbuselibrary{theorems, skins}

\usepackage{mdframed}

\title{\boldmath Signatures of kinetic gravity braiding in cosmological probes of the gravitational field
}

\newcommand{\hiclass}{\texttt{hi\_class}\xspace}

\newcommand{\gev}{\texttt{gevolution}\xspace}
\newcommand{\kev}{\texttt{k-evolution}\xspace}
\newcommand{\kgb}{\texttt{KGB-evolution}\xspace}

\author[a]{Ahmad Nouri-Zonoz \orcidlink{0009-0006-6164-8670},}
\affiliation[a]{Département de Physique Théorique and Centre for Astroparticle Physics,
    Université de Genève,
    24 quai Ernest-Ansermet,
    1211 Genève 4,
    Switzerland}
\author[b]{Farbod Hassani \orcidlink{0000-0003-2640-4460},}
\affiliation[b]{
    Institute of Theoretical Astrophysics,
    University of Oslo,
    Sem Sælands vei 13,
    0371 Oslo,
    Norway
}
\author[a,c,d]{Julian Adamek \orcidlink{0000-0002-0723-6740},}
\affiliation[c]{Institut f\"ur Teilchen- und Astrophysik, ETH Z\"urich, Wolfgang-Pauli-Strasse 27, 8093 Z\"urich, Switzerland}
\affiliation[d]{Institut f\"ur Astrophysik, Universit\"at Z\"urich,
Winterthurerstrasse 190, 8057 Z\"urich, Switzerland}
\author[e]{Emilio Bellini \orcidlink{0000-0003-4762-0795},}
\affiliation[e]{    Center for Astrophysics and Cosmology,
    University of Nova Gorica,
    Vipavska cesta 13,
    Rožna Dolina,
    SI-5000 Nova Gorica,
    Slovenia}
\author[a]{and Martin Kunz \orcidlink{0000-0002-3052-7394}\,}

\emailAdd{Ahmadreza.Nourizonoz@unige.ch}
\emailAdd{farbod.hassani@astro.uio.no}
\emailAdd{adamekj@ethz.ch}
\emailAdd{emilio.bellini@ung.si}
\emailAdd{martin.kunz@unige.ch}

\abstract{We study the observational signatures of kinetic gravity braiding (KGB) models in relativistic cosmological probes constructed along the past light cone. Using the relativistic $N$-body code \kgb, we generate light-cone outputs and compute several observables that directly probe the gravitational field, including weak gravitational lensing convergence, Shapiro time delay, the integrated Sachs–Wolfe and Rees–Sciama (ISW–RS) effects, and gravitational redshift. Full-sky maps and angular power spectra of these quantities are constructed and compared with $k$-essence models and predictions from linear perturbation theory. We find that the derivative coupling between the scalar field and the metric modifies both the amplitude and the time evolution of the gravitational potentials, producing scale-dependent deviations ranging from a few percent to tens of percent. In particular, the ISW--RS signal exhibits the largest fractional response, as the slower decay of the Weyl potential suppresses the KGB signal in the ISW-dominated regime, whereas nonlinear evolution reverses this trend at higher multipoles, producing differences of tens of percent relative to $k$-essence. Weak gravitational lensing also provides a strong complementary probe and, for the model considered here, exhibits clear deviations from the $k$-essence prediction at small scales with enhancements up to $\sim 10$--$12\%$ at multipoles $\ell \sim 10^2$--$10^3$. 
Our results show that linear perturbation theory accurately describes the large-scale behaviour, while nonlinear effects become important at smaller scales, particularly for the ISW--RS signal and, more moderately, for the convergence, and must therefore be included for reliable theoretical predictions.
}

\begin{document}
\maketitle
\flushbottom

\section{Introduction}
\label{sec:intro}
Cosmological observations probe the gravitational field through the propagation of light across an inhomogeneous universe. In the standard $\Lambda$CDM model, the evolution of the gravitational potentials is closely linked to the growth of matter perturbations and the expansion history. Extensions of gravity and dark energy models, however, can modify these potentials and therefore change the expected signals in relativistic observables. In particular, quantities such as weak gravitational lensing, the integrated Sachs–Wolfe effect, and gravitational redshift directly trace the metric perturbations along the line of sight and offer a sensitive probe of departures from standard gravity. This makes such observables particularly relevant in the context of current and upcoming galaxy surveys, where the expansion history and growth of large-scale structure are measured with increasing precision. For instance, the Dark Energy Spectroscopic Instrument (DESI) \citep{DESI:2025fii} has provided high-accuracy measurements of baryon acoustic oscillations across a wide redshift range. Together with Euclid \citep{EUCLID:2011zbd} and Stage-IV surveys such as the Nancy Grace Roman Space Telescope \citep{Akeson:2019biv} and the Vera C. Rubin Observatory \cite{LSSTScience:2009jmu}, these measurements will probe structure formation at the percent level. At this stage, achieving comparable theoretical accuracy therefore requires robust predictions for both the nonlinear evolution of matter and metric perturbations, particularly in models that extend or modify the standard $\Lambda$CDM framework.

A broad class of alternatives to the $\Lambda$CDM model arises within scalar-tensor theories of gravity. Among these, the Horndeski class \cite{Horndeski:1974wa} provides the most general scalar-tensor framework with second-order equations of motion and has become a widely used theoretical setting for studying modified gravity scenarios. The cosmological dynamics of Horndeski models at the level of linear perturbations have been implemented in Boltzmann solvers such as \hiclass \cite{Zumalacarregui:2016pph,Bellini:2019syt} and \texttt{EFT-CAMB} \cite{Hu:2013twa}. Going beyond linear theory requires numerical $N$-body simulations in order to capture the nonlinear growth of structure. In this direction, the simplest subclass of Horndeski theories, known as the $k$-essence model \citep{Armendariz-Picon:2000nqq}, has been implemented in the relativistic $N$-body code \gev \citep{Adamek:2015eda,Adamek:2016zes} under the name \kev \citep{Hassani:2019lmy,Hassani:2019wed}. More recently, some of us extended this framework to include the kinetic gravity braiding (KGB) model \citep{Deffayet:2010qz}, resulting in the \kgb code \citep{Nouri-Zonoz:2025cul}. There, we studied the impact of the braiding interaction on the clustering properties of dark energy and its interplay with the nonlinear growth of matter perturbations. In particular, we showed that the derivative coupling between the scalar field and the metric modifies the evolution of scalar perturbations, enabling dark energy fluctuations to cluster efficiently even on quasi-linear scales, including scales beyond the sound horizon where $k$-essence fluctuations are suppressed. This clustering acts as an additional source for the gravitational potentials and induces characteristic modifications in statistical quantities such as the potential and matter power spectra.

While these results establish the impact of braiding on matter clustering and the gravitational potentials, it remains important to assess how these modifications translate into observable signatures. In particular, many cosmological probes that are sensitive to the gravitational field depend on integrated effects along the line of sight and therefore require a consistent treatment on the past light cone of an observer. In this work, we therefore extend the analysis of the KGB models to the level of light-cone observables. Using the \kgb code, we construct the past light cone of a fiducial observer and compute sky maps and angular power spectra of several relativistic effects that probe the gravitational field, including weak gravitational lensing, the integrated Sachs–Wolfe and Rees--Sciama effects, gravitational redshift, and the Shapiro time delay.

The paper is organised as follows. In Section \ref{sec:theory_kgb} we review the theoretical framework of the KGB models, summarise the relevant perturbation equations, and discuss the resulting clustering behaviour of KGB in comparison with the $k$-essence limit. In Section \ref{sec:obs} we introduce the light-cone observables and compute the corresponding angular power spectra. Section \ref{sec:setup} outlines the numerical setup and the light-cone configuration in the simulations as well as the adopted cosmological parameters. In Section \ref{sec:lens_numeric} we present the resulting maps and angular power spectra and discuss the observable signatures of kinetic gravity braiding. Section \ref{sec:conclusions} summarises our conclusions.

\section{Linear analysis of the KGB dark energy model}
\label{sec:theory_kgb}
In this section, we briefly summarise the KGB model and highlight its impact on the evolution of cosmological perturbations. This section is based on linear calculations performed with \hiclass, with the goal of developing an intuitive understanding of the behaviour of perturbations in the KGB model that we are studying. The nonlinear evolution in KGB is discussed in the following sections in the context of light-cone observables. KGB belongs to the Horndeski class of scalar-tensor theories and extends $k$-essence by introducing a derivative coupling between the scalar field and the metric, which mixes the kinetic terms of the scalar field and the metric tensor and modifies the dynamics of perturbations. In this class of models, dark energy is described by a minimally coupled scalar field that has no direct coupling to matter or to the Ricci scalar and whose dynamics are governed by the Lagrangian density
\begin{equation}
\mathcal{L}_{\rm DE} = 
 G_2(\phi,X)
- G_3(\phi,X)\Box\phi\, ,
\label{eq:lagrange_de}
\end{equation}
where $X = -\tfrac{1}{2}\nabla_\mu\phi\nabla^\mu\phi$ and $\Box \phi = \nabla_\mu \nabla^\mu\phi$. The functions $G_2(\phi,X)$ and $G_3(\phi,X)$ are arbitrary functions of the scalar field and its kinetic term. The limit $G_3=0$ reduces the theory to standard $k$-essence. The resulting field equations follow from varying the total action, including the Einstein–Hilbert, scalar-field and matter contributions, and consist of the modified Einstein equations together with the scalar field equation of motion. Furthermore, we work in a perturbed FLRW spacetime in Poisson gauge,
\begin{equation}
\mathrm{d}s^2 = a^2(\tau)\left[
-e^{2\Psi}\mathrm{d}\tau^2
- 2B_i\,\mathrm{d}x^i\mathrm{d}\tau
+ \left(e^{-2\Phi}\delta_{ij}+h_{ij}\right)\mathrm{d}x^i\mathrm{d}x^j
\right],
\label{eq:metric}
\end{equation}
where $\Phi$ and $\Psi$ are the scalar potentials, $B_i$ is transverse, and $h_{ij}$ is
transverse and traceless. We define the scalar perturbation as
$\pi \equiv \delta\phi/\bar{\phi}'$ and define $\zeta \equiv \pi' + \mathcal{H}\pi - \Psi$ as in \cite{Hassani:2019lmy}. As customary, in the effective field theory (EFT) framework the background scalar field is used to define the time slicing, so models in which the background scalar field is non-monotonic are not captured by the EFT framework.

At the linear level, the effect of the KGB sector can be fully captured by the effective background energy density and pressure of the scalar field $(\bar{\rho}_\phi,\bar{P}_\phi)$, together with the EFT functions $\alpha_{\rm K}$ and $\alpha_{\rm B}$, known as \textit{kineticity} and \textit{braiding}, respectively. In this framework, the KGB sector enters the perturbation dynamics through two main ingredients:
the linearised stress-energy tensor of dark energy, which sources the metric perturbations, and
the linear equation of motion for the scalar field perturbation $\pi$. In Poisson gauge and at linear order in the metric and the scalar field, the stress-energy tensor of the scalar field reads\\
\begin{equation} 
\begin{aligned}
T^0_0 &= - \bar{\rho}_\phi + \frac{M_\textrm{Pl}{}^2}{a^2} \bigg\{\alpha_\textrm{B} \mathcal{H}(\nabla^{2}\pi) + 3 \alpha_\textrm{B} \mathcal{H}^2 \Psi - (3 \alpha_\textrm{B} + \alpha_\textrm{K}) \mathcal{H}^2 \zeta + 3 \alpha_\textrm{B} \mathcal{H} \Phi^{\prime} \\ &\hspace{3.cm}+\Big[ \alpha_\textrm{B} \mathcal{H}^{\prime}- \alpha_\textrm{B} \mathcal{H}^2 +\frac{a^2}{M_\textrm{Pl}{}^2} (\bar{\rho}_\phi + \bar{P}_\phi )\Big]3 \mathcal{H} \pi \bigg\} \, , \\[8pt] T^0_i &=-(\bar{\rho}_\phi+\bar{P}_\phi)\partial_{i}\pi +\frac{M_\textrm{Pl}{}^2 \alpha_\textrm{B}\mathcal{H}}{a^2}\partial_i \zeta  \, , \\[8pt] T^i_j &= \big(\bar{P}_\phi + \pi \bar{P}_{\phi}^{\prime}\big) \delta^i_j  \\ &\quad - \frac{M_\textrm{Pl}{}^2}{a^2}\left\{\alpha_\textrm{B} \mathcal{H}\zeta^{\prime} +\zeta \left[2 \alpha_\textrm{B} \mathcal{H}^2 + \mathcal{H}\alpha_\textrm{B}^{\prime} + \alpha_\textrm{B}\mathcal{H}^{\prime} - \frac{a^2}{M_\textrm{Pl}^2}(\bar{\rho}_\phi + \bar{P}_\phi)\right]\right\}\delta^i_j  \, . 
\end{aligned}\label{eq:DE-Tmunu} 
\end{equation} 
Here $\mathcal{H}\equiv a'/a$ denotes the conformal Hubble parameter, $M_{\rm Pl}$ is the
Planck mass, and a prime denotes a derivative with respect to conformal time
$\tau$.
  
The expressions above show how the KGB sector contributes to the perturbation dynamics through the stress-energy tensor of dark energy, which sources the metric perturbations via the Einstein field equations. The impact of braiding is controlled by the parameter $\alpha_{\rm B}$, with the limit $\alpha_{\rm B}=0$ recovering the $k$-essence case. For the numerical results presented in this section, we adopt the
$\Omega_{\rm DE}$-proportional parametrisation,
$
\alpha_i(\tau)=\hat{\alpha}_i\Omega_{\rm DE}(\tau),
$
where $\hat{\alpha}_i$ denotes the constant amplitude of the corresponding EFT function.

Although both $k$-essence and KGB are \textit{minimally} coupled scalar-tensor theories, the $G_3 \Box\phi$ term in KGB, as mentioned before, introduces  a derivative interaction that intertwines the kinetic operators of the scalar field and metric sectors. In particular, braiding generates terms proportional to $\nabla^2\pi$ in both the scalar field equation and the stress-energy tensor, which may become dominant on small scales. In the $k$-essence limit, by contrast, pressure support efficiently suppresses spatial inhomogeneities once modes enter the sound horizon, causing dark energy perturbations to decay. As shown in our previous work \cite{Nouri-Zonoz:2025cul}, the presence of this braiding-induced gradient term prevents such smoothing and allows dark energy fluctuations to grow efficiently even on small scales. The resulting clustering acts as an additional, scale-dependent source for the gravitational potentials which drives an evolution that departs from that found in $\Lambda$CDM and $k$-essence models.

It is important to note that in $k$-essence there is, to a very good approximation, a one-to-one mapping between the scalar sound speed and the efficiency of dark energy clustering: decreasing $c_s^2$ shifts the sound horizon to smaller scales and allows dark energy perturbations to grow over a wider range of modes, while increasing $c_s^2$ suppresses clustering at larger scales. In this limit, the sound speed is tightly tied to the kineticity, such that larger values of $\alpha_{\rm K}$ typically correspond to smaller $c_s^2$ and hence more prominent clustering.

\begin{figure}[t!]
    \centering
    \includegraphics[width=\linewidth]{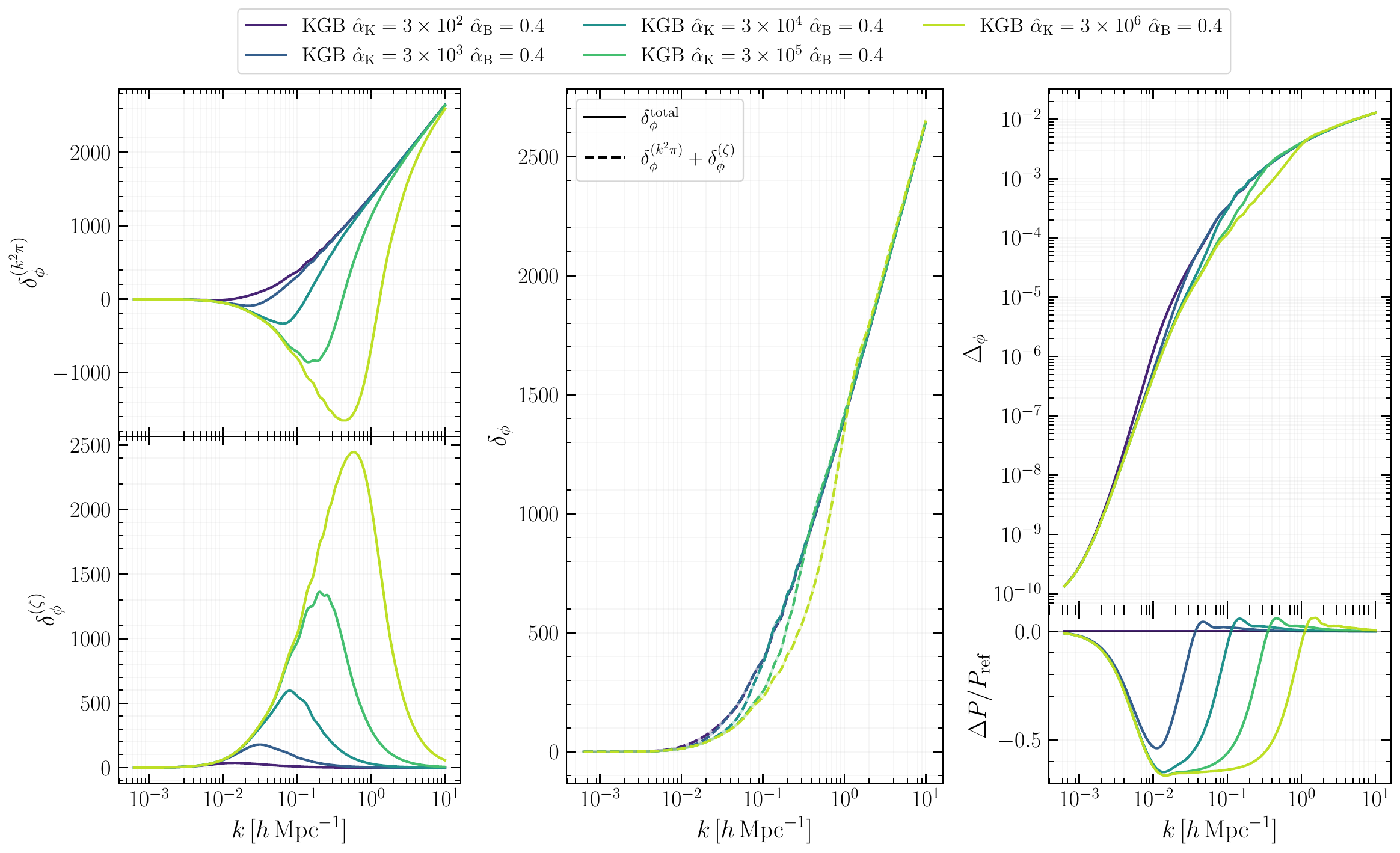}
\caption{\textit{Left:} decomposition of the KGB dark energy density contrast into its two dominant contributions defined in Eqs.~\eqref{eq:k2pi} and \eqref{eq:zetafact} (top and bottom panels respectively), for $\hat{\alpha}_{\rm B}=0.4$ and different values of $\hat{\alpha}_{\rm K}$. 
\textit{Middle:} comparison between the full density contrast $\delta_\phi$ (solid) and the sum of the two dominant contributions defined in Eqs.~\eqref{eq:k2pi} and \eqref{eq:zetafact} (dashed), showing that these terms capture the scale dependence of $\delta_\phi^{\rm total}$ over the relevant $k$-range. 
\textit{Right:} the corresponding dimensionless dark energy scalar field power spectrum $\Delta_\phi$ (top) and its relative difference with respect to the reference model with $\hat{\alpha}_{\rm K}=3\times10^{2}$ (bottom). All panels are shown at $z=0$.}
    \label{fig:three_plots}
\end{figure}

In KGB models, this one-to-one mapping between the clustering amplitude and the value of $c_s^2$ is no longer reliable, i.e.\ the correspondence is broken once the braiding parameter is active ($\alpha_{\rm B}\neq 0$). In fact, our numerical results show the opposite trend to the naive $k$-essence expectation: for $\alpha_{\rm B}>0$, decreasing $\alpha_{\rm K}$ increases the sound speed,
\begin{equation}
c_s^2(\tau)
= \frac{-\alpha_{\mathrm{B}}^2\mathcal{H}^2 +2\mathcal{H}\alpha'_{\mathrm{B}}
+2\alpha_{\mathrm{B}}\mathcal{H}'
+\frac{2a^2}{M_{\rm Pl}^2}\bigl(\bar\rho_\phi+\bar P_\phi\bigr)}
{\mathcal{H}^2\bigl(3\alpha_{\mathrm{B}}^2+2\alpha_{\mathrm{K}}\bigr)} , 
\label{eq:soundspeed}
\end{equation}
yet leads to more efficient clustering of dark energy on intermediate scales. The physical origin of this behaviour can be understood directly from the structure of the scalar field density perturbation in Eq.~\eqref{eq:DE-Tmunu}, where one can write the density contrast in Fourier space as 
\begin{equation}
\begin{aligned}
\delta_{\phi}  &=    \frac{M_\textrm{Pl}{}^2}{a^2 \bar{\rho}_\phi} \bigg\{-\alpha_\textrm{B} \mathcal{H}  k^2\pi  + 3 \alpha_\textrm{B} \mathcal{H} ^2 \Psi  - (3 \alpha_\textrm{B}  + \alpha_\textrm{K}) \mathcal{H} ^2 \zeta  + 3 \alpha_\textrm{B} \mathcal{H}  \Phi ^{\prime}
    \\&\hspace{2.2cm}+\Big[ \alpha_\textrm{B} \mathcal{H} ^{\prime}-  \alpha_\textrm{B} \mathcal{H} ^2   +\frac{a^2}{M_\textrm{Pl}{}^2} (\bar{\rho}_\phi  + \bar{P}_\phi )\Big]3 \mathcal{H} \pi \bigg\} \, .
    \label{eq:delta_rho}
\end{aligned}
\end{equation}
On the scales relevant for large-scale structure, $\delta_{\phi}$ is dominated by a competition between two contributions,
\begin{align}
\delta^{(k^2\pi)}_\phi &= -\,\frac{M_\textrm{Pl}{}^2}{a^2 \bar{\rho}_\phi}\alpha_{\rm B}\,\mathcal{H}\,k^2\pi\,, \label{eq:k2pi}\\
\delta^{(\zeta)}_\phi  &= -\,\frac{M_\textrm{Pl}{}^2}{a^2 \bar{\rho}_\phi}(3\alpha_{\rm B}+\alpha_{\rm K})\,\mathcal{H}^2\,\zeta\,,\label{eq:zetafact}
\end{align}
while the remaining terms in Eq.~\eqref{eq:delta_rho} are negligible in comparison, as will be discussed later. The resulting dark energy clustering signal is therefore controlled by the interplay between two leading contributions, which can partially cancel or reinforce it depending on scale and parameter choice. This behaviour is illustrated in Fig.~\ref{fig:three_plots}, where the two contributions defined in Eqs.~\eqref{eq:k2pi} and \eqref{eq:zetafact} dominate the decomposition of the density contrast and their sum closely tracks the full $\delta_\phi$ over the relevant range of wavenumbers.

\begin{figure}[t!]
    \centering
    \includegraphics[width=\linewidth]{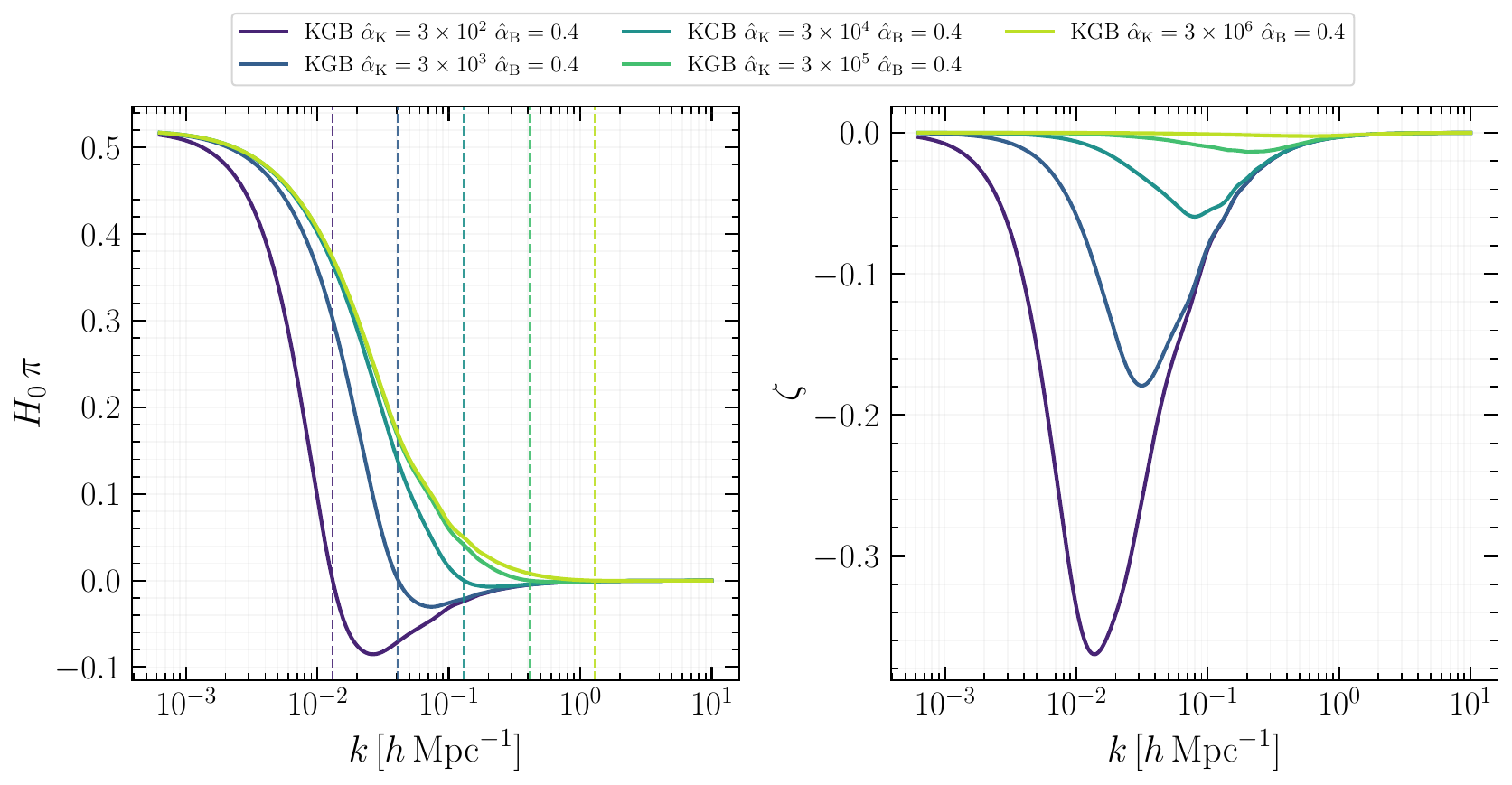}
    \caption{Scale dependence of the KGB perturbations $\pi$ (\textit{left} panel) and $\zeta$ (\textit{right} panel) at $z=0$, shown for $\hat{\alpha}_{\rm B}=0.4$ and different values of $\hat{\alpha}_{\rm K}$. Vertical dashed lines mark the wavenumber at which $H_0\pi$ crosses zero for different values of the kineticity parameter.
    }
    \label{fig:pi_zeta_alone}
\end{figure}

A further point that is clear from Fig.~\ref{fig:three_plots} is that increasing $\alpha_{\rm K}$ reduces the net density contrast on intermediate scales: although the amplitudes of the two dominant pieces become large, their sum leads to a smaller $\delta_\phi^{\rm total}$ and therefore weaker clustering compared to the cases with smaller $\alpha_{\rm K}$. This is a direct manifestation of a stronger cancellation between  Eqs.~\eqref{eq:k2pi} and \eqref{eq:zetafact} at large $\alpha_{\rm K}$, as is evident from the growing negative contribution of \eqref{eq:k2pi} and the simultaneously enhanced positive bump of \eqref{eq:zetafact} in the left panels of Fig.~\ref{fig:three_plots}. The origin of these trends can be understood by inspecting the underlying perturbations $\pi$ and $\zeta$ themselves. As shown in Fig.~\ref{fig:pi_zeta_alone}, braiding induces a zero-crossing in $\pi$ that shifts to higher $k$ as $\hat\alpha_{\rm K}$ increases. Consequently, $\pi$ remains positive down to smaller physical scales for higher $\hat\alpha_{\rm K}$, allowing the $k^2$ enhancement in $\delta_\phi^{(k^2\pi)}\propto-k^2\pi$ to generate an increasingly pronounced negative minimum before the sign reversal, as seen in Fig.~\ref{fig:three_plots}.
In contrast, the \eqref{eq:zetafact} term is controlled not only by the scale dependence of $\zeta(k)$ but also by the explicit prefactor $(3\alpha_{\rm B}+\alpha_{\rm K})$.
Therefore, although $\zeta$ itself can be larger in magnitude for smaller $\alpha_{\rm K}$, the explicit prefactor $(3\alpha_{\rm B}+\alpha_{\rm K})$  typically varies much more strongly with $\alpha_{\rm K}$ and outweighs the comparatively mild change in $\zeta(k)$. As a result, the net amplitude of the $\delta_\phi^{(\zeta)}$ contribution still increases with $\alpha_{\rm K}$. 
This explains why intermediate-scale clustering can be enhanced in KGB even when the sound speed is larger: once $\alpha_{\rm B}\neq 0$, the amplitude is set by the balance of the dominant contributions to $\delta_\phi$, rather than by $c_s^2$ alone. This is quantified in the right panel of Fig.~\ref{fig:three_plots}, where we show the dimensionless scalar field power spectrum, defined for a generic dimensionless field $X$ by the standard convention
\begin{equation}
\Delta_X(k,z) = \frac{k^3}{2\pi^2} P_X(k,z) \,.  
\end{equation}
The spectrum and its relative differences exhibit a non-trivial, scale-dependent response to changes in $\alpha_{\rm K}$.

\begin{figure}[t]
    \centering
    \includegraphics[width=\linewidth]{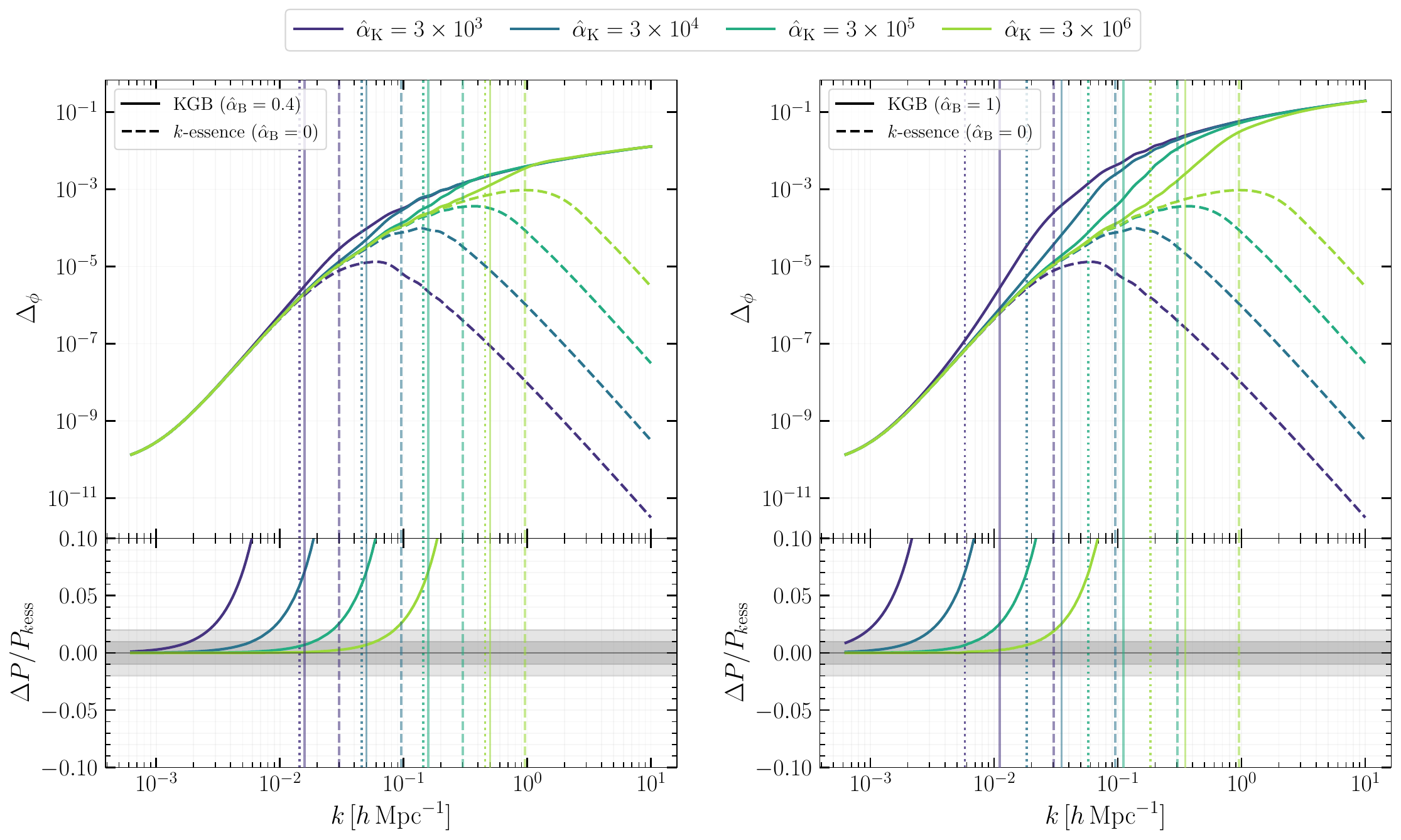}
    \caption{Dimensionless dark-energy power spectrum comparison between KGB and $k$-essence at $z=0$ for different values of the kineticity $\hat{\alpha}_{\rm K}= \{3\times10^3,\,3\times10^4,\,3\times10^5,\,3\times10^6\}$. \textit{Left:} dimensionless dark-energy power spectra in KGB (solid) with braiding $\hat{\alpha}_{\rm B}=0.4$ compared to the corresponding $k$-essence limit (dashed, $\hat{\alpha}_{\rm B}=0$). \textit{Right:} same, but for $\hat{\alpha}_{\rm B}=1$. In both columns, the lower panels show the fractional difference $\Delta P / P_{k\text{ess}}$. For each $\hat{\alpha}_{\rm K}$, the vertical solid and dashed lines mark the KGB and $k$-essence sound-horizon scales, respectively, while the vertical dotted lines indicate the KGB braiding scale.
    }

    \label{fig:DE_kgb_vs_kess}
\end{figure}

Another aspect that will be useful for our later discussion is the comparison between dark energy clustering in KGB and in its $k$-essence limit. Fig.~\ref{fig:DE_kgb_vs_kess} shows the dimensionless scalar field power spectrum $\Delta_\phi$ in KGB (solid) and in $k$-essence (dashed), for different values of the kineticity $\hat{\alpha}_{\rm K}=\{3\times10^3,\,3\times10^4,\,3\times10^5,\,3\times10^6\}$ and for two representative choices of the braiding strength, $\hat{\alpha}_{\rm B}=\{0.4,\,1\}$; all spectra shown are evaluated at $z=0$. The lower panels show the corresponding fractional difference $\Delta P/P_{k\text{ess}}$.

Two main physical scales control the behaviour of the perturbations. In the $k$-essence limit, the relevant scale is the sound horizon, beyond which scalar perturbations are efficiently smoothed by pressure support. This defines a characteristic wavenumber,
\begin{equation}
 k_{\rm s}(z)\equiv \frac{2\pi}{r_{\rm s}}  = \frac{2\pi}{\displaystyle \int_{0}^{a(z)} \frac{c_s(a)}{a^2H(a)}\,da } \, ,
\end{equation}
such that modes with $k \gtrsim k_{\rm s}$ lie inside the sound horizon and are therefore subject to efficient pressure smoothing, whereas modes with $k \lesssim k_{\rm s}$ remain outside and can cluster more effectively. In KGB models, however, the dynamics is modified due to the braiding, and this introduces an additional characteristic scale, the \emph{braiding scale} $k_{\rm B}$ \citep{Bellini:2014fua, Sawicki:2012re},
\begin{equation}
\frac{k_{\mathrm{B}}^2}{a^2 H^2} \equiv \frac{\alpha_{\rm K}+\frac{3}{2} \alpha^2_{\rm B}}{\alpha_{\mathrm{B}}^2}\Big[(1-\Omega_{\rm m})\left(1+w_\phi\right)\Big]+\frac{9}{2} \Omega_{\rm m} \, ,
\end{equation}
where $H$ is the physical Hubble
parameter, $\Omega_{\rm m} \equiv \bar{\rho}_{\rm m}/(3M_{\rm Pl}^2H^2)$ denotes the matter density fraction, and $w_\phi \equiv \bar{P}_\phi/\bar{\rho}_\phi$ is the effective equation-of-state parameter of the scalar field.
As discussed in \citep{Bellini:2014fua}, the braiding scale $k_{\rm B}$ marks the transition in the behaviour of the gravitational potential $\Phi$. For $k \ll k_{\rm B}$, the dynamics remains close to the $k$-essence limit, while for $k \gg k_{\rm B}$ the evolution of perturbations is modified, leading to a different behaviour of $\Phi$. These scales are indicated in Fig.~\ref{fig:DE_kgb_vs_kess} by vertical lines: for each value of $\hat{\alpha}_{\rm K}$, the solid and dashed lines mark the KGB and $k$-essence sound horizons, respectively, while the dotted line marks the KGB braiding scale.

Two trends are apparent. First, at fixed braiding, increasing the kineticity shifts both the sound horizon and the braiding scale to smaller physical scales. As a result, the agreement between KGB and $k$-essence extends to progressively higher $k$, reflecting the fact that a smaller sound speed delays the departure of scalar perturbations between the two models. Second, at fixed kineticity, increasing the braiding parameter shifts the braiding scale to larger physical scales. As a result, KGB clustering becomes more efficient relative to $k$-essence already on comparatively large scales. This reflects the fact that braiding counteracts the efficient smoothing of scalar perturbations that characterises the $k$-essence regime once modes enter the sound horizon, leading to earlier departures from $k$-essence behaviour.

\begin{figure}[h]
    \centering
    \includegraphics[width=\linewidth]{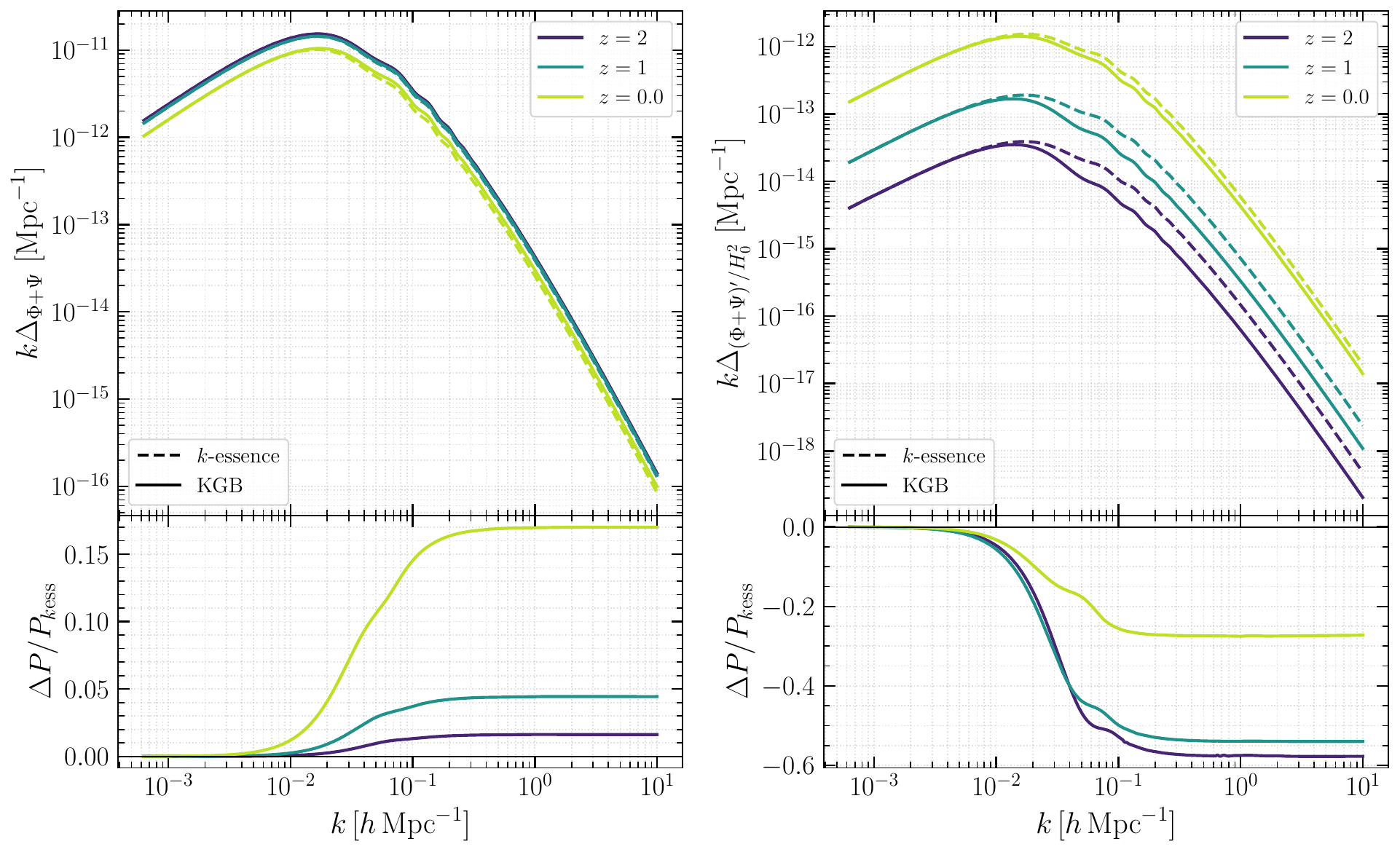}
    \caption{Scale-weighted dimensionless power spectra of the Weyl potential (\textit{Left}) and its time derivative (\textit{Right}) in KGB (solid) and $k$-essence (dashed) models for $\hat{\alpha}_{\rm K}=3\times10^{3}$ and $\hat{\alpha}_{\rm B}=0.4$ at redshifts $z=\{2,1,0\}$. The lower panels show the fractional differences.}
    \label{fig:weyl}
\end{figure}

In addition to the clustering of the scalar field itself, it is also important to examine how the scalar field dynamics affects the evolution of the gravitational potentials. A particularly relevant quantity in this respect is the Weyl potential $\Phi_{\rm W} \equiv \Phi+\Psi$, whose time derivative encodes the temporal evolution of the metric perturbations and directly enters observables such as the ISW--RS effect. In Fig.~\ref{fig:weyl} we show the power spectra of the Weyl potential and its time derivative for KGB (solid lines) and its $k$-essence limit (dashed lines) at different redshifts. The left panels display the Weyl potential power spectrum, while the right panels show the corresponding spectrum for its time derivative. The lower panels in each column present the relative difference with respect to the $k$-essence case.

The left column shows that the amplitude of the Weyl potential is enhanced in KGB relative to $k$-essence on intermediate and small scales. This behaviour reflects the more efficient clustering of dark energy in the presence of braiding. While dark energy perturbations in $k$-essence are efficiently smoothed inside the sound horizon, the kinetic braiding of the metric and the scalar field in KGB prevents this suppression and allows dark energy fluctuations to remain relevant on smaller scales. As a consequence, dark energy contributes more effectively to the sourcing of the gravitational potentials, leading to a larger amplitude of the Weyl potential compared to the $k$-essence limit, as clearly visible in the upper-left panel of Fig.~\ref{fig:weyl}. The relative differences shown in the lower-left panel confirm that this enhancement grows towards smaller scales where the impact of dark energy clustering becomes increasingly significant. Since these results are based on linear perturbation theory, they can be trusted only up to $k \sim 0.1\,h/\mathrm{Mpc}$; on smaller scales nonlinear effects become important and modify the behaviour, as shown in Fig.~3 of \cite{Nouri-Zonoz:2025cul}.

A different trend can be seen in the right column, which shows the power spectrum of the time derivative of the Weyl potential. In this case, the KGB curves lie below their $k$-essence counterparts, indicating that the time variation of the gravitational potentials is smaller in the presence of braiding. This behaviour can be understood from the competition between two physical effects. On the one hand, the accelerated background expansion driven by dark energy causes the gravitational potentials to decay at late times. On the other hand, the enhanced clustering of dark energy in the KGB model continuously contributes to the sourcing of the gravitational potentials, partially compensating the decay induced by the background expansion. As a result, the Weyl potential evolves more slowly in time and consequently the amplitude of $P_{(\Phi+\Psi)'}$ is suppressed relative to the $k$-essence case, as seen in the upper-right panel of Fig.~\ref{fig:weyl}. The lower-right panel further illustrates that this suppression can reach the level of tens of percent on intermediate and small scales. We note, however, that at late times the decay of the Weyl potential persists towards smaller scales within linear theory, while the growth associated with the nonlinear Rees--Sciama effect is not captured. In the nonlinear evolution, this effect leads to a transition from decaying to growing potentials, which occurs at larger scales in KGB compared to $k$-essence, as we will see in Section \ref{sec:lens_numeric}.

These results show that while braiding enhances the amplitude of the gravitational potentials through more efficient dark energy clustering, it simultaneously reduces their rate of temporal evolution. This interplay between enhanced sourcing and background-driven decay plays an important role for observables sensitive to the time variation of the gravitational field, such as the integrated Sachs–Wolfe effect that we will discuss in the following sections.

\section{Relativistic observables on the past light cone}
\label{sec:obs}

Cosmological information is encoded in the properties of light received from distant sources. As photons travel from their emission point to the observer, their trajectories and energies are continuously affected by inhomogeneities in the gravitational field which are sourced by different species such as matter, radiation and dark energy via the Poisson equation. These perturbations leave characteristic imprints on observables such as apparent positions, redshifts, and arrival times of photons. Importantly, several of these effects depend directly on the gravitational potentials and their evolution along the line of sight, which makes them sensitive probes of the growth of structure and the nature of dark energy. In this section, we focus on observables that arise from the propagation of light in a perturbed FLRW spacetime.

The cumulative impact of the metric perturbations on the photon propagation can be expressed as a mapping between the unperturbed source position and the observed position inferred from redshift and angular measurements. At leading order, this mapping naturally separates into distinct contributions associated with redshift perturbations, transverse deflection of light rays and line-of-sight time delays. As a result, one can write the apparent comoving 3D position of the source as \citep{Hassani:2020buk,PhysRevD.84.043516, PhysRevD.84.063505,PhysRevD.80.083514, 10.1093/mnras/sty3206, PhysRevD.101.023512}
\begin{equation}
\begin{aligned}
    \boldsymbol{s} = \chi_s \boldsymbol{n}
    &-\underbrace{\int_0^{\chi_s}\left(\chi_s-\chi\right) \nabla_{\perp}(\Phi+\Psi) \mathrm{d} \chi}_{\text{weak gravitational lensing}}-\underbrace{\boldsymbol{n} \int_0^{\chi_s}(\Phi+\Psi) \mathrm{d} \chi}_{\text{Shapiro time delay}} \\
&+ \frac{1+\bar{z}}{H_s(\bar{z})}\;\bigg[\underbrace{\boldsymbol{n} \cdot\left(\boldsymbol{v}_s-\boldsymbol{v}_o\right)}_{\text {Doppler }}+\underbrace{\Psi_o-\Psi_s}_{\begin{array}{c}
\substack{\text{gravitational}}\\[-1ex]
\substack{\text{redshift}}
\end{array}}-\underbrace{\int_0^{\chi_s} \frac{\partial(\Psi+\Phi)}{\partial \tau} d \chi}_{\text {ISW--RS effect }}\bigg]\boldsymbol{n} \,.
\end{aligned}
\label{eq:3Dposition}
\end{equation}
In the above expression, $\boldsymbol{s}$ is the apparent comoving position of the source as inferred by the observer. The unit vector $\boldsymbol{n}$ is the unperturbed line-of-sight direction from the observer to the source, while $\chi_s$ is the background comoving distance to the source corresponding to the unperturbed redshift $\bar z$. The integration variable $\chi$ represents the comoving radial distance along the unperturbed photon trajectory.

The second term describes the transverse deflection of light rays due to weak gravitational lensing. It is sourced by the gradient of the Weyl potential $\Phi +\Psi$ projected orthogonally to the line of sight, where $\nabla_\perp \equiv (\mathbf{I}-\boldsymbol{n}\boldsymbol{n}^T)\nabla$. The prefactor $\chi_s -\chi$ encodes the efficiency of a deflection occurring at  comoving distance $\chi$ in altering the apparent angular position of a source located at $\chi_s$. This term is responsible for distortions in the observed shapes and sizes of background sources and forms the basis of weak lensing observables such as cosmic shear and convergence.

The third term corresponds to Shapiro time delay, which arises because photons propagating through gravitational potentials experience perturbations in their travel time. When this is expressed in terms of an inferred comoving position, this effect manifests as an additional displacement along the line of sight which is proportional to the line-of-sight integral of the gravitational potentials. 

The second line of Eq. \eqref{eq:3Dposition} accounts for the radial displacement of the source position induced by redshift perturbations $\delta z = z - \bar z$. This contribution arises because, in a perturbed spacetime, the observed redshift $z$ differs from its background value $\bar z$, leading to a mismatch between the redshift-inferred comoving distance and the true background distance. Therefore, $\delta z$ parametrises the deviation from the one-to-one mapping between redshift and comoving distance. As discussed extensively in the literature (e.g. \citep{PhysRevD.84.063505,PhysRevD.84.043516}), the redshift perturbation itself receives contributions from several physical effects. These include, at leading order, local terms such as Doppler and gravitational redshift contributions associated with the source and observer, as well as an integrated contribution sourced by the time evolution of the gravitational potentials along the photon trajectory. The latter gives rise to the integrated Sachs–Wolfe effect and its nonlinear extension, the Rees–Sciama effect, and provides a direct probe of the dynamics of the gravitational field on cosmological scales.

One should also note that the Doppler contribution, through its direct connection to the peculiar velocity field, carries significant cosmological information and is in principle considered a sensitive probe of the properties of dark energy via the growth rate of structure. In this work, however, we only study those observables that are generated by the gravitational potentials $\Phi$ and $\Psi$. Since most of the effects generated by these potentials enter observables through line-of-sight integrals, they encode the cumulative influence of the gravitational field along the photon path. Because of this integrated nature, such effects do not require the presence of visible matter at every location and remain measurable even across large underdensed regions such as voids where the impact of dark energy is expected to be particularly significant.

In the following, we will discuss and study each of the terms present in Eq. \eqref{eq:3Dposition} separately.

\subsection{Weak gravitational lensing}
As we mentioned, the transverse displacement of the photon trajectory due to weak lensing $\boldsymbol{\Delta}_{\perp}^{\rm WL}$ is given by
\begin{equation}
\boldsymbol{\Delta}_{\perp}^{\rm WL}(\chi_s) = 
-\int_0^{\chi_s}\left(\chi_s-\chi\right) \nabla_{\perp}(\Phi+\Psi) \mathrm{d} \chi \;.
\end{equation}
In the weak lensing regime relevant on cosmological scales, the deflection angles are small. The transverse comoving displacement at the source plane can therefore be related to the observed angular deflection through the small-angle approximation,
$\Delta_{\perp}^{\rm WL}\simeq\chi_s\boldsymbol{\alpha}$.
Furthermore, throughout this work we adopt the Born approximation, in which the Weyl potential and its transverse gradient are evaluated along the unperturbed photon trajectory rather than along the true deflected geodesic. Under these approximations, the weak lensing deflection angle is given by \citep{Bartelmann:1999yn}
\begin{equation}
\boldsymbol{\alpha}^{\rm WL}(\boldsymbol{n},\chi_s) \equiv \delta\boldsymbol{\theta}=
-\int_0^{\chi_s} \frac{\chi_s-\chi}{\chi_s}\nabla_{\perp}(\Phi+\Psi)\mathrm{d}\chi \; .
\end{equation}
Transverse spatial gradients can then be expressed in terms of angular derivatives on the unit sphere through $\nabla_\perp=\chi^{-1}\hat{\nabla}$, such that the deflection angle can be written as the gradient of a scalar lensing potential $\psi$,
\begin{equation}
\boldsymbol{\alpha}^{\rm WL}(\boldsymbol{n}) = \hat\nabla \psi(\boldsymbol{n},\chi_s),
\qquad
\psi(\boldsymbol{n}, \chi_s) =
-\int_0^{\chi_s} \frac{\chi_s-\chi}{\chi_s\chi}(\Phi+\Psi)\mathrm{d}\chi \; .
\label{eq:defAngle}
\end{equation}

The deflection angle $\boldsymbol{\alpha}$ encodes the shift of individual light rays. However, since this field varies across the sky, neighbouring rays are deflected by different amounts. As a result, extended sources are not only displaced but also distorted on the observer’s sky. Weak lensing observables therefore arise from the differential deflection of neighbouring light rays and are fully characterised by spatial derivatives of the deflection field $\boldsymbol{\alpha}$. The same reasoning applies to the impact of weak lensing on observed galaxy clustering. Galaxy number counts are measured within fixed solid angles on the observer's sky. Spatial variations of the deflection field can modify the mapping between a fixed observed solid angle and the corresponding region on the source plane. Consequently, weak lensing can change the number of galaxies observed within that solid angle.

To make this explicit, let $\boldsymbol{\theta}$ be the angular position on the sky. In the presence of lensing, the mapping between the unlensed and lensed directions is 
\begin{equation}
    \boldsymbol{\theta}' = \boldsymbol{\theta} + \boldsymbol{\alpha}(\boldsymbol{\theta}) \, .
\end{equation}
For two neighbouring rays separated by an infinitesimal vector $\mathrm{d}\boldsymbol{\theta}$, the separation after lensing to linear order is
\begin{equation}
    \mathrm{d}{\theta}'_a = A_{ab}\mathrm{d}{\theta}_b\, ,
\end{equation}
with the Jacobian matrix 
\begin{equation}
    A_{ab} \equiv \partial \theta'_a/\partial \theta_b = \delta_{ab} + \hat\nabla_a \hat\nabla_b \psi \, .
\end{equation}
In two dimensions, the Jacobian can be uniquely decomposed into an isotropic component, a symmetric traceless component, and an antisymmetric component
\begin{equation}
A_{ab} = (1-\kappa)\delta_{ab} - \gamma_{ab} - \omega\epsilon_{ab}\, .
\end{equation}
Here $\kappa$ is the convergence, $\gamma_{ab}$ is the shear tensor, and $\omega$ represents a local rotation of images.
The rotation term arises only at second order in the lensing distortion and is therefore subdominant compared with the convergence and shear, which are first-order quantities (see e.g.\ \citep{Lepori:2020ifz,Magi:2026xcy}). In what follows, we neglect $\omega$.
The remaining quantities are given by
\begin{align}
\label{eq:kappa}
\kappa &= -\tfrac{1}{2}\hat\nabla^2\psi \, , \\
\gamma_{ab} &= -(\hat\nabla_a\hat\nabla_b\psi - \tfrac{1}{2}\delta_{ab}\hat\nabla^2\psi) \, .   
\end{align}
The convergence $\kappa$ controls the trace of the deformation and therefore measures the local isotropic magnification of images; a positive (negative) convergence increases (decreases) the apparent angular size of a source while preserving its shape. In contrast, the shear $\gamma_{ab}$, being traceless,  describes an anisotropic deformation that stretches images along one direction and compresses them along the orthogonal one, without changing their area. The shear is therefore responsible for the characteristic elliptic distortions and coherent alignments of galaxy images observed in weak lensing surveys \citep{Mandelbaum:2018,Alsing:2015zca}. 
These properties motivate the use of weak lensing as a powerful observational probe. Following the first detections of cosmic shear \citep{vanWaerbeke:2000rm} and later measurements of lensing magnification probing the convergence field \citep{Schmidt_2012}, weak lensing observables have reached a level of precision that allows tight constraints on dark energy and modified gravity models \citep{PhysRevLett.91.141302, Amendola:2007rr,Hannestad:2006as,SpurioMancini:2018apc}. It is useful to recall that the shear field can be decomposed into rotationally invariant E- and B-mode components. For lensing sourced by scalar metric perturbations, only E-modes are generated at leading order, while B-modes are of the same order as image rotation. In this limit, and neglecting observational systematics and shape noise, the convergence and shear fields carry equivalent cosmological information \citep{Kohlinger:2017sxk,Becker:2012qe,Hassani:2020buk}. Their angular power spectra are related through
\begin{equation}
C_\ell^{\gamma \rm E} =
\frac{(\ell+2)!}{\ell^2(\ell+1)^2(\ell-2)!}\,
C_\ell^{\kappa}\, ,
\qquad
C_\ell^{\gamma \rm B} \propto C_\ell^{\omega} \ll C_\ell^\kappa \, .
\end{equation}
In the following, we therefore focus on the convergence power spectrum.

To quantify its statistical properties, we characterise weak lensing at the level of two-point statistics. Since the convergence $\kappa(\boldsymbol{n})$ and lensing potential $\psi(\boldsymbol{n})$ are scalar fields on the sphere, they can be decomposed in terms of spherical harmonic functions

\begin{equation}\label{eq:sphericalDec}
\psi(\boldsymbol{n}) = \sum_{\ell m} \psi_{\ell m} Y_{\ell m}(\boldsymbol{n})\, , 
\qquad
\kappa(\boldsymbol{n}) = \sum_{\ell m} \kappa_{\ell m} Y_{\ell m}(\boldsymbol{n}) \, ,
 \end{equation}
with coefficients
\begin{equation}
\psi_{\ell m}=\int d\Omega_{\boldsymbol{n}}Y^*_{\ell m}(\boldsymbol{n})\psi(\boldsymbol{n})\, ,
\qquad
\kappa_{\ell m}=\int d\Omega_{\boldsymbol{n}}Y^*_{\ell m}(\boldsymbol{n})\kappa(\boldsymbol{n})\, .
\label{eq:coeff}
\end{equation}
Statistical isotropy implies
\begin{equation}
\langle \psi_{\ell m}\psi^*_{\ell' m'} \rangle
= \delta_{\ell\ell'}\delta_{mm'}\,C_\ell^\psi\, ,
\qquad
\langle \kappa_{\ell m}\kappa^*_{\ell' m'} \rangle
= \delta_{\ell\ell'}\delta_{mm'}\,C_\ell^\kappa \, .
\label{eq:statiso}
\end{equation}
These relations define the angular power spectra
$C_\ell^\psi$ and $C_\ell^\kappa$ of the lensing potential and
convergence, respectively. Using Eq. \eqref{eq:kappa} and the fact that $\hat{\nabla}^2 Y_{\ell m} = -\ell (\ell + 1) Y_{\ell m}$ one finds the relation between harmonic coefficients and therefore between the spectra as
\begin{equation}
    \kappa_{\ell m} = \frac{1}{2}\ell(\ell + 1) \psi_{\ell m}\, , 
    \qquad
        C_{\ell}^\kappa = \frac{1}{4}\ell^2 (\ell + 1)^2 C_\ell^\psi \, .
        \label{eq:relation}
\end{equation}

Thus, computing $C_\ell^\kappa$ reduces to computing the spectrum of the projected lensing potential $\psi$ and all the information in the convergence two-point function is contained in the angular power spectrum of the projected lensing potential $\psi$.

Starting from Eq.~\eqref{eq:defAngle}, the lensing potential is written as a line-of-sight projection of the Weyl potential. Using the spherical-harmonic decomposition of $\psi(\boldsymbol n)$ given in Eq.~\eqref{eq:sphericalDec}, and decomposing the Weyl potential into Fourier modes, one can relate the harmonic coefficients $\psi_{\ell m}$ to the unequal-time power spectrum of the Weyl potential. This leads to the expression
\begin{align}
C_\ell^\psi(\chi_s)
&= \langle \psi_{\ell m}(\chi_s)\psi_{\ell m}^*(\chi_s) \rangle \nonumber\\
&= \frac{2}{\pi}
\int_0^{\chi_s} d\chi\, W_\psi(\chi,\chi_s)
\int_0^{\chi_s} d\chi'\, W_\psi(\chi',\chi_s)
\int_0^\infty k^2\, dk\,
j_\ell(k\chi)\,j_\ell(k\chi')\,
P_{\Phi+\Psi}(k;\chi,\chi') \, ,
\label{eq:Clpsi_exact_weyl}
\end{align}
where $
W_\psi(\chi,\chi_s) \equiv (\chi_s-\chi)/\chi_s\chi
$
is the lensing kernel and $P_{\Phi+\Psi}(k;\chi,\chi')$ denotes the unequal-time power spectrum of the Weyl potential. Using the Limber approximation \citep{1954ApJ...119..655L}, and neglecting unequal-time correlations, this expression can be evaluated straightforwardly. Combining this with Eq.~\eqref{eq:relation}, one obtains the convergence angular power spectrum
\begin{equation}
C_{\ell}^{\kappa}(\chi_s)
\simeq
\left(\frac{\ell(\ell+1)}{2}\right)^{2}
\int_{0}^{\chi_{s}}\!d\chi\;
\frac{W_{\psi}(\chi,\chi_{s})^{2}}{\chi^{2}}\,
P_{\Phi+\Psi}\!\left(k=\frac{\ell+1/2}{\chi},\,\chi\right)\,.
\label{eq:Clkappa_limber}
\end{equation}

\subsection{Shapiro time delay}

From the second term in Eq.~\eqref{eq:3Dposition}, we obtain the Shapiro time delay 
\begin{equation}
    \Delta\tau = -\int_0^{\chi_s}(\Phi+\Psi) \mathrm{d} \chi\,.
\end{equation}
This effect arises from the fact that gravitational potentials perturb the local passage of time, so that photons propagating through overdense or underdense regions experience a shift in their travel time. Originally proposed as a test of general relativity in \citep{PhysRevLett.13.789}, it provides a direct probe of the gravitational potential integrated along the line of sight.

In contrast to weak lensing, which is sourced by transverse gradients of the potentials and leads to angular distortions of images, the Shapiro time delay depends directly on the projected Weyl potential itself. As a result, it does not deflect photon trajectories, but instead induces radial distortions that modify the inferred distance to the source. In geometric terms, surfaces of constant emission time are distorted along the line of sight, rather than across it.

Because the Shapiro delay depends on the potential rather than its derivatives, its statistical properties differ significantly from those of weak lensing. In particular, the absence of spatial gradients suppresses the contribution from small-scale modes, such that the signal is dominated by large-scale fluctuations of the gravitational potential. Consequently, most of the power is expected at low multipoles, in contrast to weak lensing observables which receive significant contributions from smaller scales.

A useful illustration of the Shapiro time delay is provided by its effect on the cosmic microwave background (CMB). The last scattering surface already contains intrinsic perturbations that generate the primary CMB anisotropies. The Shapiro time delay constitutes an additional propagation effect on top of these intrinsic fluctuations. Spatial variations in the integrated Weyl potential perturb the travel times of CMB photons, causing photons observed in different directions to correspond to different radial positions on the already perturbed last scattering surface. In this sense, the time delay produces a further line-of-sight remapping of the CMB source surface, with a characteristic radial displacement of order $\mathcal{O}(\mathrm{Mpc})$ \citep{Hu:2001yq}.

In the context of CMB anisotropies, it is useful to distinguish between contributions that directly perturb the observed photon energy at linear order and secondary propagation effects. Among the potential-dependent effects considered here, the ISW effect and gravitational redshift, which are discussed in more detail in the following subsections, belong to the first category, whereas weak gravitational lensing and the Shapiro time delay act mainly as secondary propagation effects.

Although this secondary contribution is difficult to isolate observationally, it affects CMB temperature and polarisation anisotropies and must be included in sufficiently precise analyses to avoid systematic biases \citep{Hu:2001yq}. More recently, it has been shown that dedicated estimators can, in principle, reconstruct maps of the Shapiro time delay from CMB data, with a potentially detectable signal on the largest angular scales \citep{Li:2019qkp}. Furthermore, since the time delay is sourced by the gravitational potential itself, it provides a sensitive probe of large-scale potential fluctuations.

We now quantify the statistical properties of the Shapiro time delay at the level of two-point statistics. The corresponding angular power spectrum can be written as
\begin{equation}
C_\ell^{\Delta\tau}(\chi_s)
=
\frac{2}{\pi}
\int_{0}^{\chi_s} d\chi
\int_{0}^{\chi_s} d\chi'\;
\int_{0}^{\infty} dk\,k^2\;
j_\ell(k\chi)\,j_\ell(k\chi')\;
P_{\Phi+\Psi}(k;\chi,\chi') \, ,
\label{eq:Cl_deltatau_exact}
\end{equation}
where $P_{\Phi+\Psi}$ denotes the unequal-time power spectrum of the Weyl potential. Using the Limber approximation and neglecting unequal-time correlations, this expression reduces to
\begin{equation}
C_\ell^{\Delta\tau}(\chi_s)
\simeq
\int_{0}^{\chi_s}\!d\chi\;
\frac{1}{\chi^{2}}\;
P_{\Phi+\Psi}\!\left(k=\frac{\ell+1/2}{\chi},\,\chi\right) \, .
\label{eq:Cl_deltatau_limber}
\end{equation}
The absence of a geometric weighting kernel in Eq.~\eqref{eq:Cl_deltatau_limber} indicates that, unlike weak lensing, the Shapiro time delay does not include an additional source-dependent efficiency factor such as $(\chi_s-\chi)/(\chi_s\chi)$. While the signal is still integrated along the line of sight, its weighting is determined solely by the projection and the evolution of the Weyl potential power spectrum. Consequently, it probes the integrated large-scale distribution of the Weyl potential across the light cone without preferentially selecting specific distances.

\subsection{Integrated Sachs--Wolfe and Rees--Sciama effect}

From the last term in Eq.~\eqref{eq:3Dposition}, the ISW--RS contribution is
\begin{equation}
I_{\rm ISW-RS}(\boldsymbol{n}, \chi_s)
\equiv
-\int_0^{\chi_s}
\frac{\partial(\Phi+\Psi)}{\partial \tau}
\, d \chi \, .
\end{equation}


The ISW--RS effect describes the change in photon energy induced by time-varying gravitational potentials along the line of sight \citep{Sachs:1967er}. For a generic source, we characterise this effect through the dimensionless redshift perturbation
\begin{equation}
\Theta(\mathbf{n},\chi_s)
\equiv
-\frac{1}{1+\bar z}
\left.\delta z\right|_{\rm ISW-RS}
=
\int_0^{\chi_s} d\chi\,
\frac{\partial(\Phi+\Psi)}{\partial\tau}.
\label{eq:theta_isw_rs}
\end{equation}
For CMB photons, this quantity is conventionally interpreted as the fractional temperature perturbation, $\Theta=\Delta T/\bar T$. Unlike the Shapiro time delay, which depends on the value of the potential itself, the ISW--RS contribution is sensitive to its time variation and therefore vanishes if the potentials remain constant along the line of sight.

A useful way to interpret this effect is to note that it isolates departures from a purely matter-dominated evolution. In such a regime, the gravitational potentials are approximately constant at linear order, and therefore no ISW contribution is generated. Any nonzero contribution therefore directly reflects a change in the dynamical evolution of the Universe, making this observable particularly sensitive to late-time acceleration or modifications of gravity.

Beyond this large-scale behaviour, additional contributions arise once the growth of structure becomes nonlinear. In this regime, the evolution of gravitational potentials is driven by the local dynamics of collapsing and evolving structures, leading to the Rees--Sciama effect \citep{Rees:1968zza}. This contribution is physically distinct from the linear ISW effect, as it is sourced by the nonlinear growth of potential wells rather than their decay. 

At the level of the angular power spectrum, however, the separation between the linear ISW and nonlinear RS contributions is not direct. Although the ISW--RS field can be conceptually decomposed into a part sourced by the linear decay of the gravitational potentials and a part sourced by their nonlinear evolution, the observed spectrum of the total field $\Theta$ contains the corresponding auto- and cross-correlations,
\begin{equation}
C_\ell^\Theta
=
C_\ell^{\rm lin}
+
C_\ell^{\rm nl}
+
2 C_\ell^{\rm cross}.
\end{equation}
As a result, the signs of the underlying linear and nonlinear contributions are not directly reflected in the angular power spectrum. Moreover, the projection
along the line of sight mixes contributions from different scales and times, which further smooths the transition between the linear and nonlinear regimes. The RS effect therefore appears as an additional small-scale contribution to $C_\ell^\Theta$, where nonlinear structure formation becomes important
(see e.g.\ \citep{Cai:2010hx}).

Taken together, the ISW and RS effects probe the time evolution of the gravitational potential across both linear and nonlinear regimes. Because of this, they provide complementary information to weak lensing and the Shapiro time delay, which depend respectively on spatial gradients and line-of-sight integrals of the potential. In practice, isolating the ISW signal from CMB data alone is challenging, and it is typically accessed through cross-correlations with tracers of the large-scale structure \citep{Adamek:2019vko,Cabass:2015xfa, Khosravi:2015boa, Beck:2018owr}.

We now quantify the statistical properties of this effect. Proceeding as in the previous cases, the angular power spectrum can be written as
\begin{equation}
C_\ell^{\Theta}(\chi_s)
=
\frac{2}{\pi}
\int_{0}^{\chi_s} d\chi
\int_{0}^{\chi_s} d\chi'\;
\int_{0}^{\infty} dk\,k^2\;
j_\ell(k\chi)\,j_\ell(k\chi')\;
P_{(\Phi+\Psi)'}(k;\chi,\chi') \, ,
\end{equation}
where $P_{(\Phi+\Psi)'}$ denotes the unequal-time power spectrum of the time derivative of the Weyl potential. Using the Limber approximation and neglecting unequal-time correlations, this reduces to
\begin{equation}
C_\ell^{\Theta}(\chi_s)
\simeq
\int_{0}^{\chi_s}\!d\chi\;
\frac{1}{\chi^{2}}\;
P_{(\Phi+\Psi)'}\!\left(k=\frac{\ell+1/2}{\chi},\,\chi\right)
\, .
\end{equation}

\subsection{Gravitational redshift}

From Eq.~\eqref{eq:3Dposition}, the gravitational redshift contribution arises
from the difference of the gravitational potential between the source and the
observer,
\begin{equation}
\delta z_{\rm grav} = (1+\bar{z}) (\Psi_o - \Psi_s) \, .
\end{equation}

This term originates from the energy shift experienced by photons when climbing out of or falling into gravitational potential wells. In contrast
to the other relativistic contributions discussed above, it is a purely local effect: it depends only on the values of the potential at the endpoints of the photon trajectory and does not involve any accumulation along the line of sight. For this reason, it isolates the local gravitational environment of the source and observer, rather than the integrated properties of the large-scale structure. Early theoretical work showed that gravitational redshift effects could reach observable amplitudes in massive galaxy clusters \citep{1995A&A...301....6C}. This prediction was later confirmed observationally through statistical analyses of large galaxy samples \citep{Wojtak:2011ia}. In clusters and massive dark matter halos, galaxies located at different positions within the gravitational potential acquire small systematic shifts in their observed redshifts. Since the signal from individual systems is weak, it is generally extracted statistically by stacking large samples of structures. Measurements of this type have provided evidence for gravitational redshift on megaparsec scales and have enabled tests of gravity in the weak-field regime
\citep{Sadeh:2014rya,Jimeno:2014xma,Zhu:2019cix}.

Gravitational redshift is not restricted to galaxies within clusters, but can also be probed statistically through the cross-correlation of distinct galaxy populations, such as bright and faint galaxies. Since bright galaxies typically reside in more massive halos, they occupy deeper gravitational potentials and acquire, on average, a larger gravitational redshift contribution than faint galaxies. This relative radial displacement breaks the front--back symmetry of bright--faint pairs along the line of sight and generates odd multipoles in their cross-correlation function, most notably the dipole.

An initial detection of this asymmetric signal was reported using BOSS galaxies, with much of the measured signal arising from separations below approximately $10\,h^{-1}\mathrm{Mpc}$ \citep{Alam:2017}. Forecasts and modelling studies indicate that upcoming surveys may measure this dipole over both linear and nonlinear scales \citep{Saga:2021,SobralBlanco:2022,Lepori:2024,Dam:2025}. Its interpretation requires other contributions to the asymmetry, including Doppler, wide-angle, light-cone, and nonlinear effects, to be modelled consistently.

These observational approaches do not measure the gravitational potential field directly. Instead, the gravitational redshift contribution enters the observed galaxy distribution together with density, velocity, lensing, and other relativistic terms \citep{PhysRevD.84.063505}, and is inferred through its correlation with the spatial distribution of galaxies or other matter tracers. Consequently, such measurements are sensitive to a density--potential cross-spectrum $P_{\delta\Psi}$, rather than directly to the potential auto-spectrum $P_{\Psi}$. In the present work, we instead consider the angular auto-spectrum of the gravitational redshift field itself. This provides a direct characterisation of the local potential contribution and facilitates comparison with the other gravitational observables discussed here.

Since the observer term contributes only to the monopole, the angular fluctuations considered here are entirely determined by the potential at the source position. The corresponding angular auto-spectrum is therefore obtained by projecting the local potential power spectrum $P_{\Psi}$ onto a spherical shell at comoving distance $\chi_s$,
\begin{equation}
C^{\delta z_{\rm grav}}_\ell(\chi_s)
=
(1+\bar z_s)^2
\frac{2}{\pi}
\int_0^\infty dk\, k^2
P_{\Psi}(k,\chi_s) j_\ell^2(k\chi_s) .
\label{eq:Cl_grav_exact}
\end{equation}

Unlike the previous cases, this expression contains no line-of-sight integral. Instead, it projects the local gravitational potential power spectrum onto a thin spherical shell at the source distance.

\section{Simulation setup and light-cone configuration}
\label{sec:setup}
We use the \kgb code to construct and analyse past light cones for a fiducial observer. The simulations are performed with the GPU-accelerated version of \kgb, which builds on the GPU implementation and parallelisation framework introduced in \cite{Adamek:2026vof}.
For the sake of comparison, we adopt a light-cone setup similar to that of 
\cite{Hassani:2020buk}. Our simulations evolve $N_{\rm pcl}=3072^3$ dark matter particles in a cubic box of comoving side length $L=4032~{\rm Mpc}/h$, while the fields are evolved on a fixed mesh with $N_{\rm grid}=3072$ points per dimension. The observer is placed at the corner of the simulation box, at coordinates $(x,y,z)=(0,0,0)$, and two distinct light cones are recorded: a full-sky light cone extending to a comoving distance $\chi_s=2015~{\rm Mpc}/h$, corresponding to $z\simeq 0.86$, and a second light cone with an opening half-angle of $25^\circ$ reaching $\chi_s=4690~{\rm Mpc}/h$, corresponding to $z\simeq 3.6$. Moreover, we adopt the cosmological parameters summarised in Table \ref{table:cosmoparams} for all simulations, where the background evolution of dark energy is specified by the Chevallier–Polarski–Linder (CPL) parametrisation \citep{Chevallier:2000qy, Linder:2002et}. We further explore $k$-essence and KGB models by varying the EFT functions 
$\alpha_{\rm K}$ and $\alpha_{\rm B}$. For each of these models we consider two cases by constructing a small grid in the EFT parameter space. 
For $k$-essence we adopt 
$(\hat{\alpha}_{\rm K},\hat{\alpha}_{\rm B}) = (3\times10^{3},\,0)$ and 
$(\hat{\alpha}_{\rm K},\hat{\alpha}_{\rm B}) = (3\times10^{6},\,0)$, 
while for KGB we consider 
$(\hat{\alpha}_{\rm K},\hat{\alpha}_{\rm B}) = (3\times10^{3},\,0.4)$ and 
$(\hat{\alpha}_{\rm K},\hat{\alpha}_{\rm B}) = (3\times10^{6},\,0.4)$. 
In all cases, we use the commonly used \texttt{propto\_omega} parametrisation, in which each EFT 
function scales with the dark energy density fraction according to
\begin{equation}
    \alpha_i(\tau) = \hat{\alpha}_i\, \Omega_{\rm DE}(\tau)\, .
\end{equation}

\begin{table}[h]
 \caption{Cosmological parameters used in the simulations}
  \centering
  \begin{tabular}{|*{9}{|c}||}
    \hline
    \rowcolor{crisp}
    $\Omega_{\text{cdm}}$ 
      & $\Omega_\text{DE}$ 
      & $\Omega_\text{b}$ 
      & $\Omega_\text{g}$ 
      & $w_0$ 
      & $w_a$ 
      & $h$ 
      & $A_s$ 
      & $n_s$ \\
    \hline\hline
    0.2638 
      & 0.6878 
      & 0.04828 
      & $5.418\times10^{-5}$ 
      & $-0.90$ 
      & 0 
      & 0.6755 
      & $2.215\times10^{-9}$ 
      & 0.9619 \\
    \hline
  \end{tabular}
  \label{table:cosmoparams}
\end{table}

\section{Numerical results from simulation}
\label{sec:lens_numeric}
\begin{figure}
  \centering
  \includegraphics[width=0.93\textwidth]{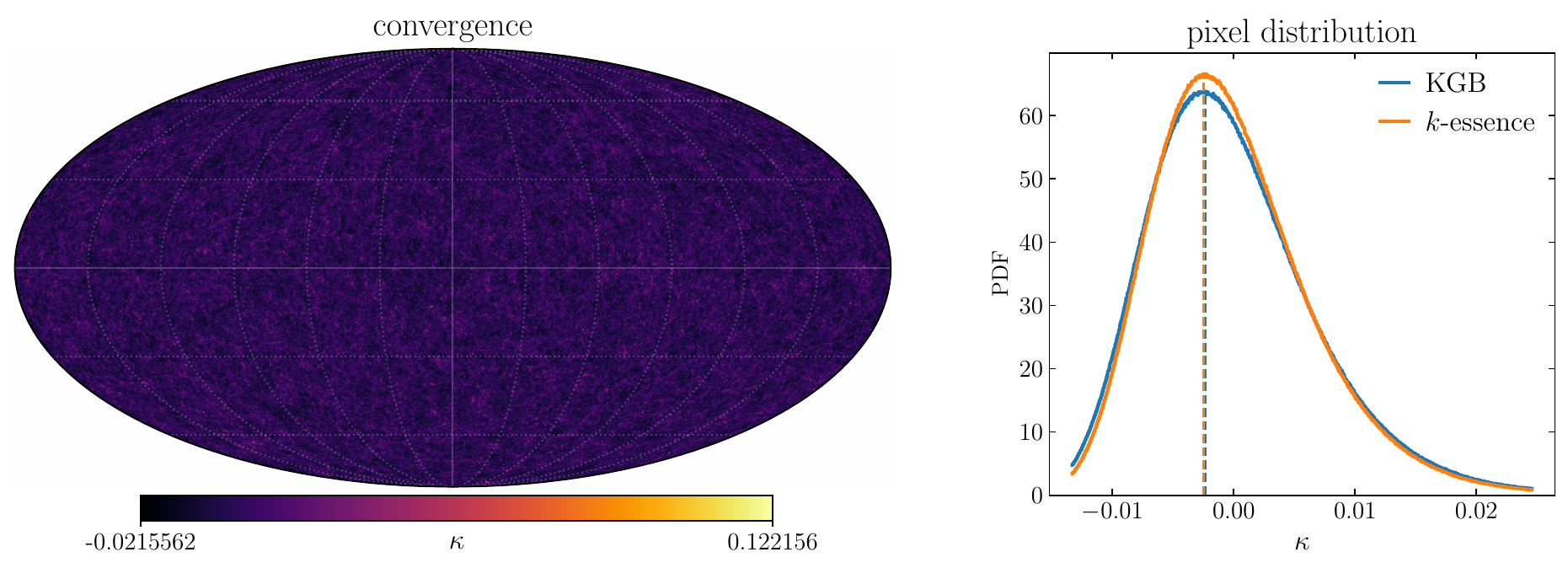}\par\vspace{0.0em}
  \includegraphics[width=0.93\textwidth]{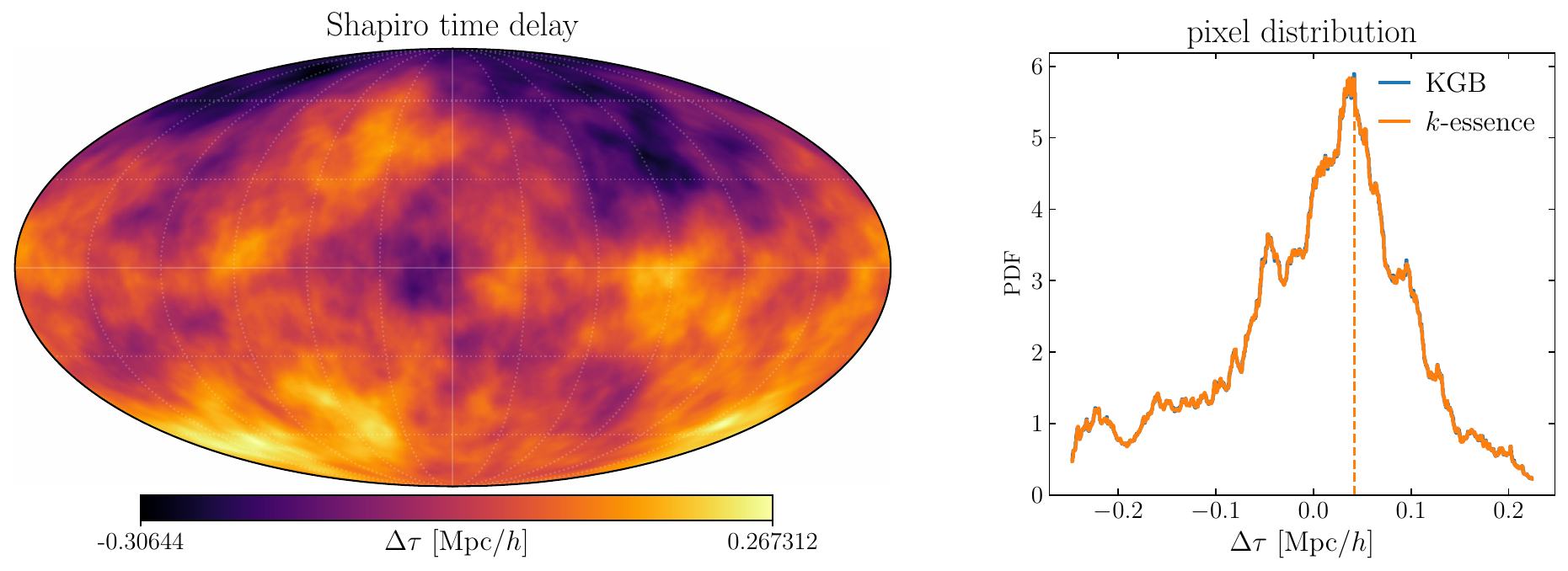}\par\vspace{0.0em}
  \includegraphics[width=0.93\textwidth]{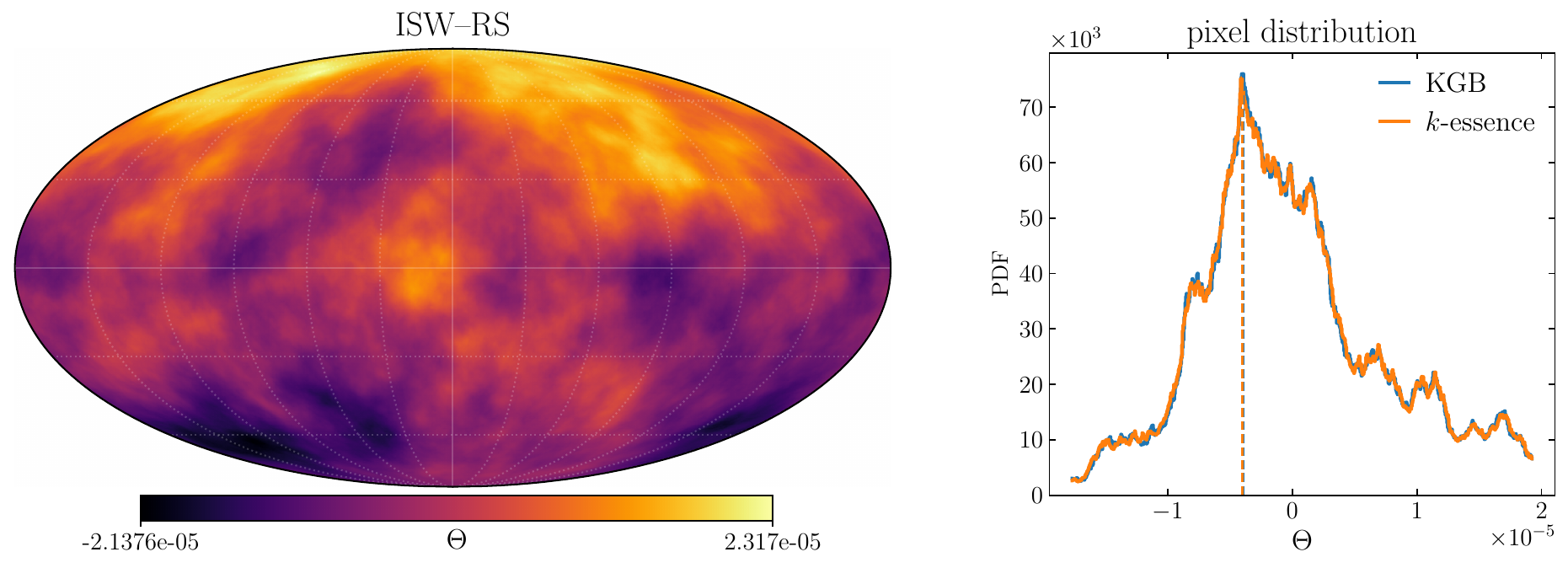}\par\vspace{0.0em}
  \includegraphics[width=0.93\textwidth]{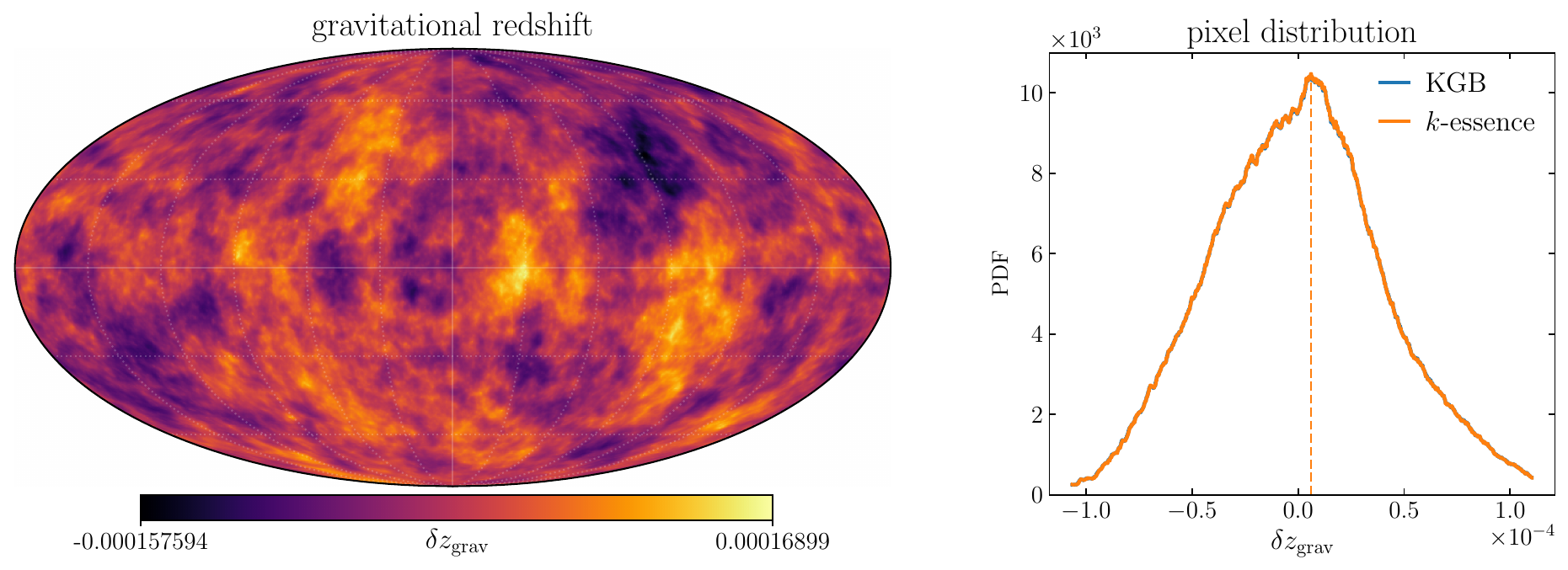}
\caption{Full-sky maps at $z=0.86$ for the KGB model with $(\hat{\alpha}_{\rm K},\hat{\alpha}_{\rm B})=(3\times10^{3},\,0.4)$: (top to bottom) convergence $\kappa$, Shapiro time delay, ISW--RS field $\Theta$, and gravitational redshift. The right-hand histograms show the corresponding pixel-value probability density functions for both KGB and the $k$-essence model with $(\hat{\alpha}_{\rm K},\hat{\alpha}_{\rm B})=(3\times10^{3},0)$.}
  \label{fig:sky_maps}
\end{figure}

In this section we present the numerical results from \kgb on the past light cone. Fig.~\ref{fig:sky_maps} shows full-sky maps at $z = 0.86$ in the KGB model $(\hat\alpha_{\rm K},\hat\alpha_{\rm B})=(3\times10^{3},0.4)$ for the different observables considered in this work, generated with the \texttt{lcmap} post-processing tool, together with the corresponding pixel-value histograms comparing the KGB model with the $k$-essence case $(\hat\alpha_{\rm K},\hat\alpha_{\rm B})=(3\times10^{3},0)$. Details of the light-cone construction in the simulations and the \texttt{lcmap} analysis pipeline are given in Appendix~\ref{app:lc}. The sky maps provide an overview of the characteristic angular morphology and typical amplitudes of each signal, while the histograms illustrate how the distribution of values changes across models. In the remainder of this section we study each observable in turn and quantify the model dependence in harmonic space through the angular power spectra, with particular emphasis on the impact of the dark energy parameters $\hat{\alpha}_{\rm K}$ and $\hat{\alpha}_{\rm B}$.

For the angular power spectra presented below, the convergence range of the absolute spectra depends on the observable, while the relative KGB--$k$-essence differences remain robust up to $\ell\simeq1000$; see Table~\ref{tab:convergence_limits} in Appendix~\ref{app:conv_test}.

\subsection*{Weak lensing convergence}
We begin with the weak lensing convergence, which directly probes the projected Weyl potential along the line of sight. Since braiding can enhance the clustering of scalar perturbations on sub-horizon scales, we expect a corresponding amplification of the gravitational potentials and therefore of the lensing signal.

\begin{figure}[h]
    \centering
    \includegraphics[width=\linewidth]{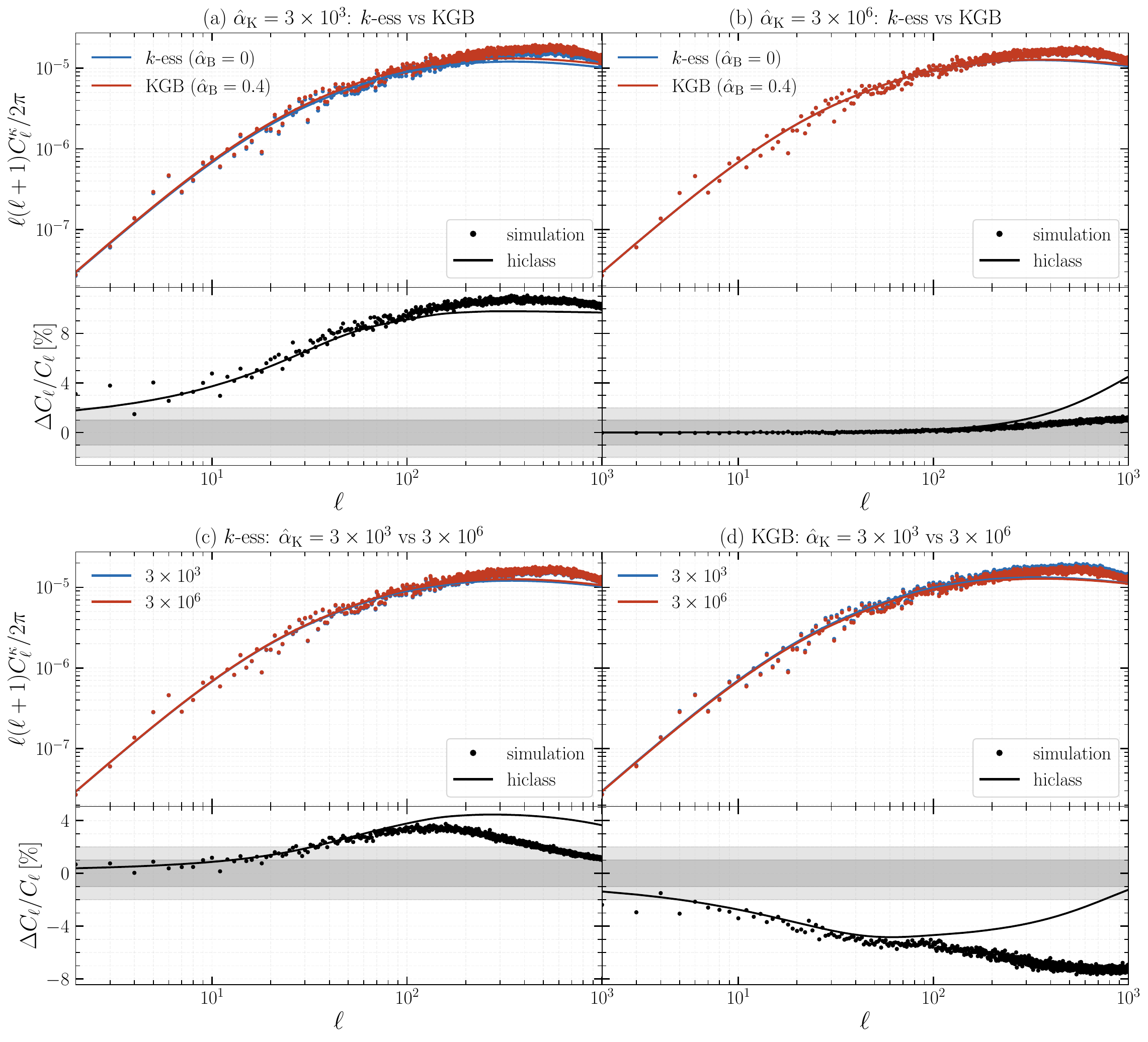}
    \caption{Angular power spectrum of the weak lensing convergence $C_\ell^\kappa$ at source redshift $z=0.86$, obtained from the simulations and from linear theory computed with \hiclass. Panels (a) and (b) compare the $k$-essence ($\hat\alpha_{\rm B}=0$) and KGB ($\hat\alpha_{\rm B}=0.4$) models for two values of the kineticity parameter, $\hat\alpha_{\rm K}=3\times10^{3}$ and $\hat\alpha_{\rm K}=3\times10^{6}$, respectively. Panels (c) and (d) show the dependence on $\hat\alpha_{\rm K}$ within the $k$-essence model ($\hat\alpha_{\rm B}=0$) and the KGB model ($\hat\alpha_{\rm B}=0.4$), respectively, comparing $\hat\alpha_{\rm K}=3\times10^{3}$ and $\hat\alpha_{\rm K}=3\times10^{6}$ in each case. In each panel the lower sub-plot shows the fractional difference with respect to the chosen reference model. The grey bands indicate the $1\%$ and $2\%$ fractional difference ranges.}
    \label{fig:Cl_kappa}
\end{figure}

Fig.~\ref{fig:Cl_kappa} shows the angular power spectrum of the weak lensing convergence computed from the light-cone output of the simulation, together with the corresponding linear-theory prediction from \hiclass. The four panels illustrate how the signal changes when varying the dark energy model and its parameters. The top row compares $k$-essence and KGB for fixed values of $\hat\alpha_{\rm K}$, while the bottom row isolates the dependence on $\hat\alpha_{\rm K}$ within each model separately. In each case we also show the fractional difference relative to the chosen reference model. 

In panel (a) we compare the $k$-essence model ($\hat\alpha_{\rm B} =0$) and the KGB model ($\hat\alpha_{\rm B} = 0.4$) for $\hat\alpha_{\rm K} = 3 \times10^3$. The KGB case exhibits a larger convergence power spectrum across the full multipole range. While the two models agree closely on large angular scales, the deviation increases progressively towards higher $\ell$, reaching nearly $12\%$ at $\ell \sim 10^{2}\text{--}10^{3}$. This behaviour is also clearly visible in the lower sub-panel, where the fractional difference grows monotonically with multipole. The scale dependence of this enhancement is consistent with the modified perturbation dynamics induced by braiding. Higher multipoles probe smaller physical scales along the line of sight, where the braiding interaction more efficiently boosts structure growth. Since the weak lensing convergence is determined by the projected Weyl potential, the increased clustering in the KGB model translates directly into an amplified Weyl power spectrum and therefore a higher amplitude of $C_\ell^\kappa$ compared to the $k$-essence case.

In panel (b) we repeat the same comparison between $k$-essence and KGB, but now for a larger value of the kineticity, $\hat\alpha_{\rm K} = 3 \times 10^6$. In this case the two models remain much closer to each other across the multipole range of interest, with only a small fractional difference visible at small scales in linear theory, as shown in the sub-panel. This behaviour is consistent with our discussion presented in Section \ref{sec:theory_kgb}, where we showed that increasing the kineticity parameter shifts the scale at which the KGB dark-energy power spectrum departs from the $k$-essence limit towards smaller scales.
In other words, for larger values of kineticity the two models exhibit nearly identical clustering properties over a wider range of wavenumbers, and the characteristic effects of braiding become important only at higher multipoles.

Panel (c) focuses on the dependence of the convergence spectrum on the kineticity parameter within the $k$-essence model. Here we compare $\hat\alpha_{\rm K} = 3 \times 10^3$ and $\hat\alpha_{\rm K} = 3 \times 10^6$, which correspond to sound speeds of $c_s^2 = 10^{-4}$ and $c_s^2 = 10^{-7}$, respectively. Again, as we discussed in Section \ref{sec:theory_kgb}, in $k$-essence there exists a close connection between the sound speed of dark energy and the clustering properties of dark energy: decreasing the sound speed shifts the sound horizon to smaller scales and allows dark energy perturbations to cluster more efficiently. Consequently, the convergence power spectrum shows a higher amplitude for larger $\hat{\alpha}_{\rm K}$, particularly at intermediate and high multipoles where smaller physical scales contribute more strongly to the lensing signal. This behaviour is consistent with the well-known result that a smaller sound speed in $k$-essence models leads to stronger dark energy clustering and correspondingly larger gravitational potentials \citep{Hassani:2020buk}.

Panel (d) shows the analogous comparison within the KGB model with $\hat{\alpha}_{\rm B} = 0.4$. In contrast to the $k$-essence case, the dependence on $\hat{\alpha}_{\rm K}$ now follows the opposite trend: the model with the smaller value $\hat{\alpha}_{\rm K} = 3 \times 10^{3}$ produces a larger convergence signal than the model with $\hat{\alpha}_{\rm K} = 3 \times 10^{6}$ at intermediate scales. This behaviour can be understood from the discussion in Section \ref{sec:theory_kgb}  and in particular from the decomposition of the dark energy density perturbation illustrated in Fig. \ref{fig:three_plots}. There we showed that in the presence of braiding the clustering amplitude is determined by the interplay between two dominant contributions to the density contrast, proportional to $k^{2}\pi$ and $\zeta$, whose relative importance depends on $\hat{\alpha}_{\rm K}$. For a fixed value of $\hat{\alpha}_{\rm B}$, increasing $\hat{\alpha}_{\rm K}$ enhances both terms but also strengthens the cancellation between them on intermediate scales, leading to a smaller net dark energy density contrast. In linear theory this behaviour already produces a reduction of the clustering amplitude at intermediate scales before the signal eventually grows again at sufficiently small scales. In the simulations, however, the nonlinear evolution tends to continue suppressing the clustering for larger $\hat{\alpha}_{\rm K}$, resulting in an overall decrease of the convergence amplitude across the multipole range shown in Fig. \ref{fig:Cl_kappa}. 

\subsection*{Shapiro time delay and gravitational redshift}
\begin{figure}[h]
    \centering
    \includegraphics[width=\linewidth]{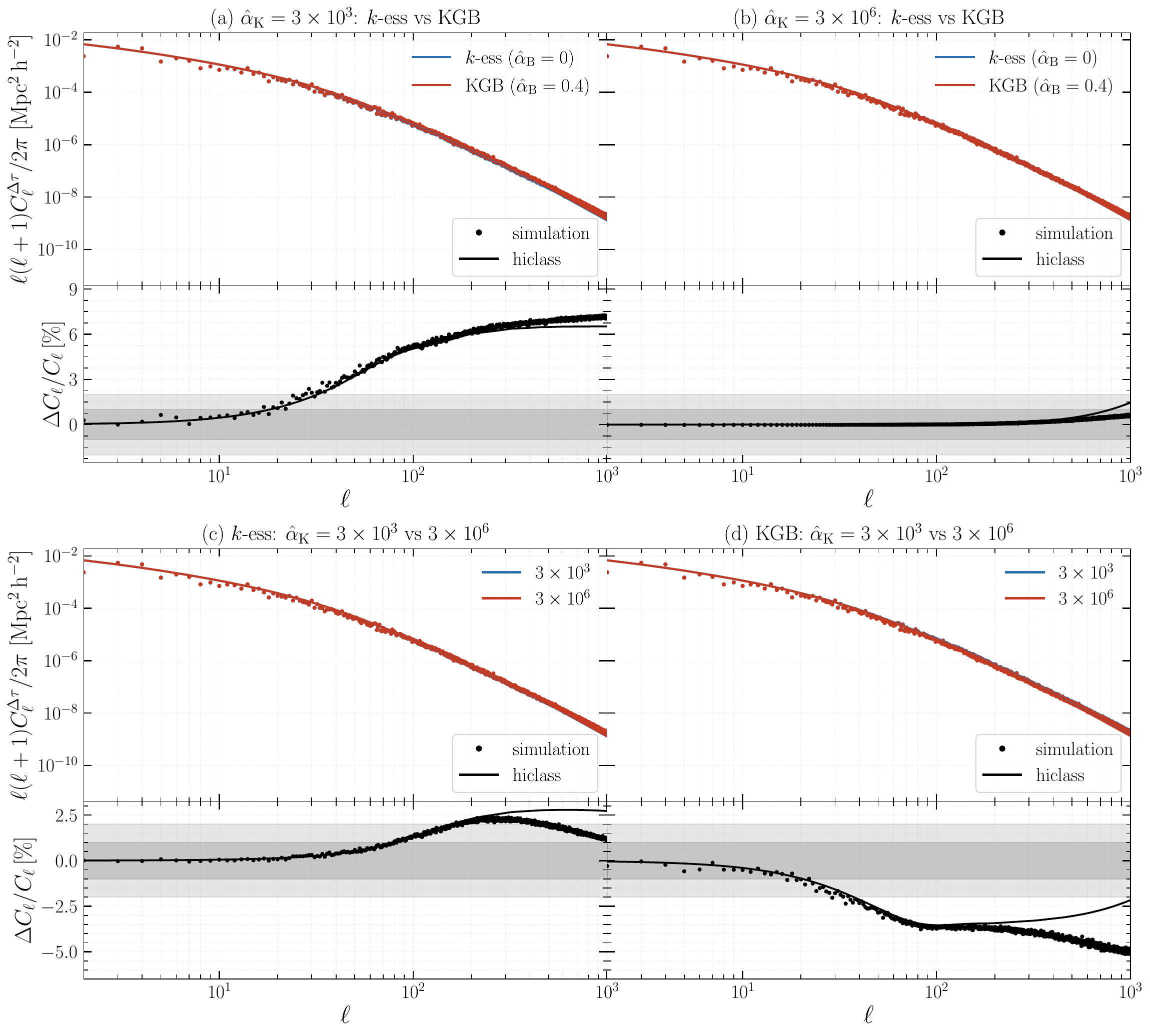}
    \caption{Angular power spectrum of the Shapiro time delay at source redshift $z=0.86$, obtained from the simulations and from linear theory computed with \hiclass. The panel layout and model choices follow Fig.~\ref{fig:Cl_kappa}: panels (a,b) compare $k$-essence ($\hat\alpha_{\rm B}=0$) and KGB ($\hat\alpha_{\rm B}=0.4$) for $\hat\alpha_{\rm K}=3\times10^{3}$ and $\hat\alpha_{\rm K}=3\times10^{6}$, while panels (c,d) show the dependence on $\hat\alpha_{\rm K}$ within each model separately. The lower panels show the fractional difference with respect to the chosen reference model. }
    \label{fig:Cl_deltatau}
\end{figure}

\begin{figure}[h]
    \centering
    \includegraphics[width=\linewidth]{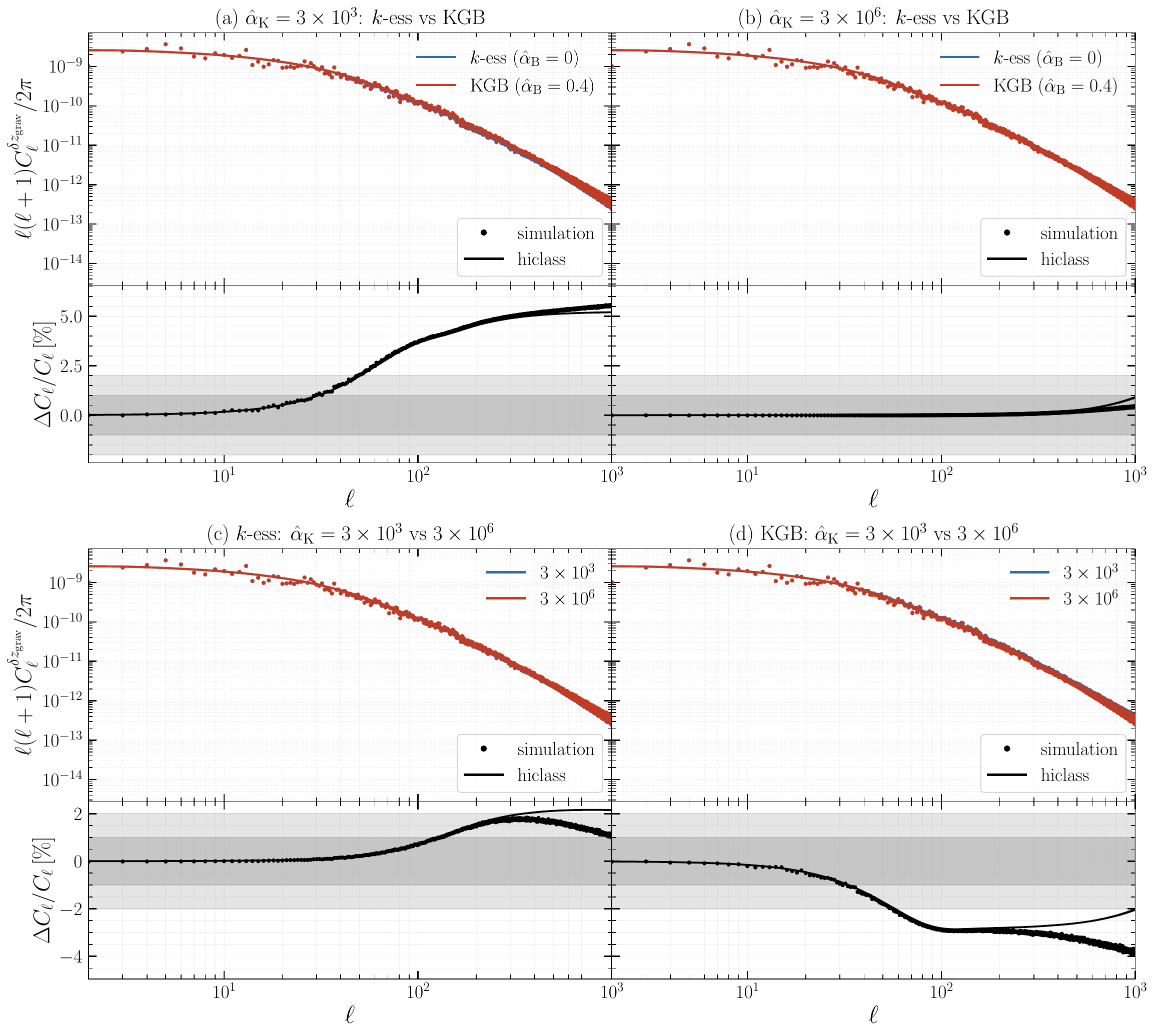}
    \caption{Angular power spectrum of the gravitational redshift at source redshift $z=0.86$, obtained from the simulations and from linear theory computed with \hiclass. The panel layout follows Fig.~\ref{fig:Cl_kappa}: panels (a,b) compare the $k$-essence ($\hat\alpha_{\rm B}=0$) and KGB ($\hat\alpha_{\rm B}=0.4$) models for $\hat\alpha_{\rm K}=3\times10^{3}$ and $\hat\alpha_{\rm K}=3\times10^{6}$, respectively, while panels (c,d) illustrate the dependence on $\hat\alpha_{\rm K}$ within each model separately. The lower panels show the fractional difference with respect to the chosen reference model.}
    \label{fig:Cl_deltaz}
\end{figure}

We next turn to the Shapiro time delay and gravitational redshift, shown in Figs.~\ref{fig:Cl_deltatau} and \ref{fig:Cl_deltaz}. Both observables probe the gravitational field, but they depend on different combinations of the metric potentials and have different geometrical origins. The Shapiro time delay is sourced by the Weyl potential $\Phi+\Psi$ integrated along the line of sight, whereas gravitational redshift is a local endpoint effect determined by the difference in the
potential $\Psi$ between the source and the observer. Their sensitivity to the dark energy model therefore differs both in amplitude and in scale dependence.

The deviations between KGB and $k$-essence are noticeably smaller than in the convergence signal. While weak lensing exhibits differences at the level of $\sim 12\%$ on intermediate and small scales, the corresponding effects in the Shapiro time delay and gravitational redshift remain typically at the few-percent level across the multipole range considered. This indicates that, for the models studied here, these observables are less sensitive to the braiding interaction than weak lensing, although they probe complementary aspects of the gravitational field.

The weaker response of these observables can be understood from the different way in which the metric potentials enter their angular power spectra. The Shapiro time delay depends on the line-of-sight projection of the Weyl-potential power spectrum $P_{\Phi+\Psi}$, as shown in Eq.~\eqref{eq:Cl_deltatau_limber}. Since it involves no transverse spatial derivatives, its signal is weighted more strongly towards large scales, where the differences between KGB and $k$-essence are comparatively small for the choices of $\hat{\alpha}$ parameters considered here. At smaller scales, additional contributions enter through the projection of the Weyl potential power spectrum, leading to a scale-dependent enhancement of the signal. This behaviour is visible in panel (a) of Fig.~\ref{fig:Cl_deltatau}, where the fractional differences show a gradual increase with multipoles, reaching $\sim 7 \%$ at $\ell \gtrsim 100$ and flattening afterwards. 

Gravitational redshift has a different dependence on the gravitational field. It is sourced by the local potential $\Psi$ at the source position, after removing the observer contribution, and its angular power spectrum is therefore determined by the thin-shell projection of $P_{\Psi}(k, z_s)$, as given in Eq.~\eqref{eq:Cl_grav_exact}. Unlike the Shapiro
time delay, it contains no line-of-sight integral and directly probes the gravitational potential on the source shell.

Therefore, although Shapiro time delay and gravitational redshift probe different aspects of the gravitational potential, they are less sensitive to braiding than weak lensing. Their main role is therefore complementary when combined with more sensitive observables.

\subsection*{Integrated Sachs–Wolfe and Rees–Sciama effect}
\begin{figure}[h]
    \centering
    \includegraphics[width=\linewidth]{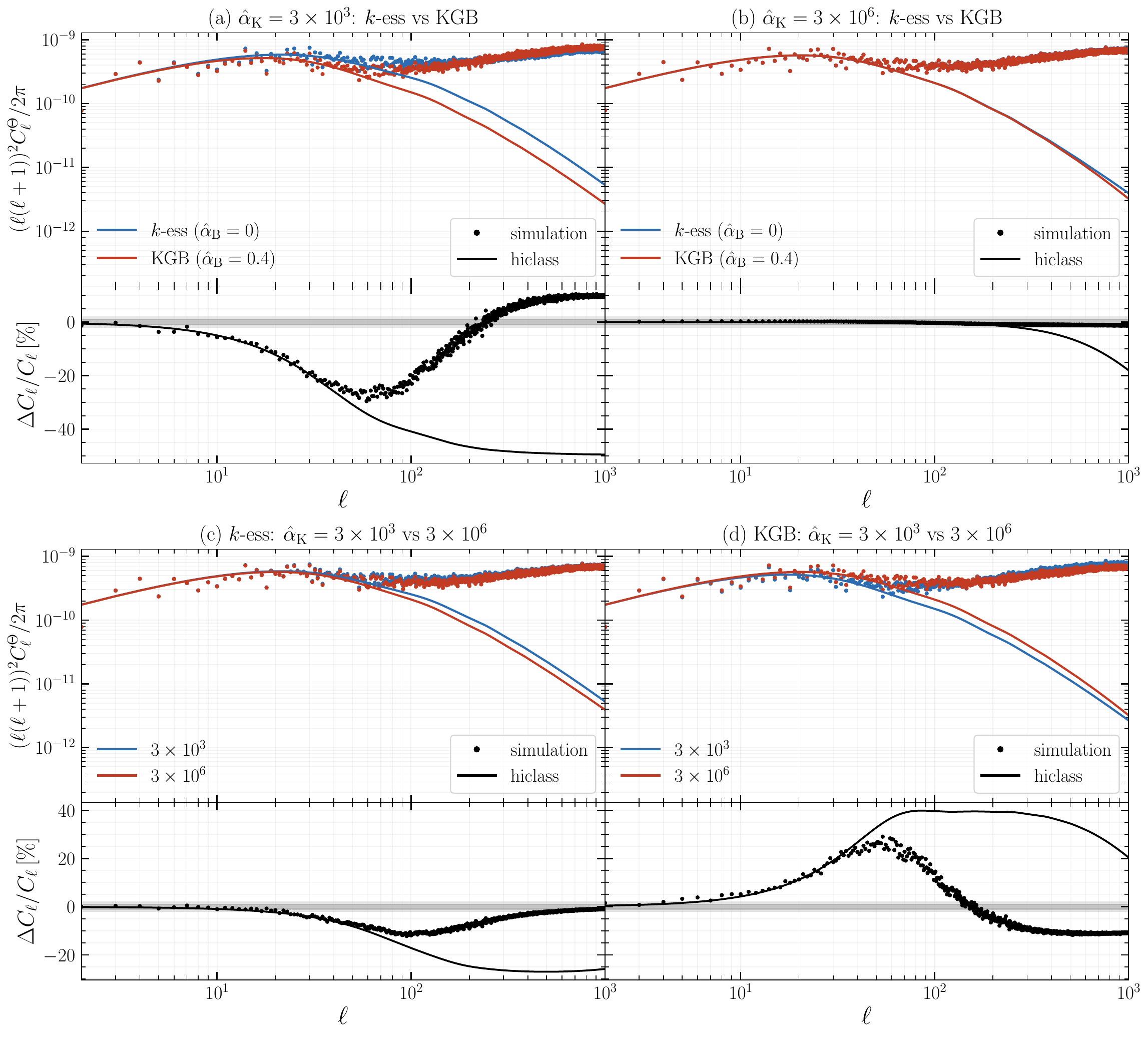}
   \caption{Angular power spectrum of the ISW--RS signal at source redshift $z=0.86$, obtained from the simulations and from linear theory computed with \hiclass. The panel layout follows Fig.~\ref{fig:Cl_kappa}: panels (a,b) compare the $k$-essence ($\hat\alpha_{\rm B}=0$) and KGB ($\hat\alpha_{\rm B}=0.4$) models for $\hat\alpha_{\rm K}=3\times10^{3}$ and $\hat\alpha_{\rm K}=3\times10^{6}$, respectively, while panels (c,d) illustrate the dependence on $\hat\alpha_{\rm K}$ within each model separately. The lower panels show the fractional difference with respect to the chosen reference model.}
    \label{fig:Cl_isw}
\end{figure}

The ISW--RS signal shown in Fig. \ref{fig:Cl_isw} exhibits a behaviour distinct from the previous observables. In contrast to the previous observables, this effect is sourced by the time derivative of the Weyl potential and therefore directly probes the temporal evolution of the gravitational field rather than its amplitude. The behaviour of the ISW--RS spectrum can be understood from the results presented in Section \ref{sec:theory_kgb} (see Fig.~\ref{fig:weyl}), where we showed that although the Weyl potential itself is enhanced in the KGB model, its time derivative is suppressed due to the slower evolution of the gravitational potentials. As a result, the ISW--RS signal exhibits a lower amplitude in the KGB case compared to $k$-essence over a wide range of multipoles, as seen in Fig. \ref{fig:Cl_isw}.  At higher multipoles, nonlinear effects become important and the signal transitions from the linear ISW regime to the Rees--Sciama (RS) regime. In this regime, the linear predictions from \hiclass start to deviate from the simulation results, reflecting the additional time variation of the gravitational potentials induced by nonlinear structure formation. This behaviour indicates that, while the linear ISW contribution is suppressed in the KGB model due to the slower evolution of the potentials, the nonlinear RS contribution is enhanced as a result of stronger clustering.

Because the ISW--RS signal is difficult to measure directly, it is often probed
indirectly through cross-correlations with other tracers of large-scale structure. Motivated by this, we consider its cross-power spectrum with the weak lensing convergence. Combining these observables allows one to relate the clustering of matter to the time evolution of the gravitational potentials in a single statistic.

\begin{figure}[h]
    \centering
    \includegraphics[width=\linewidth]{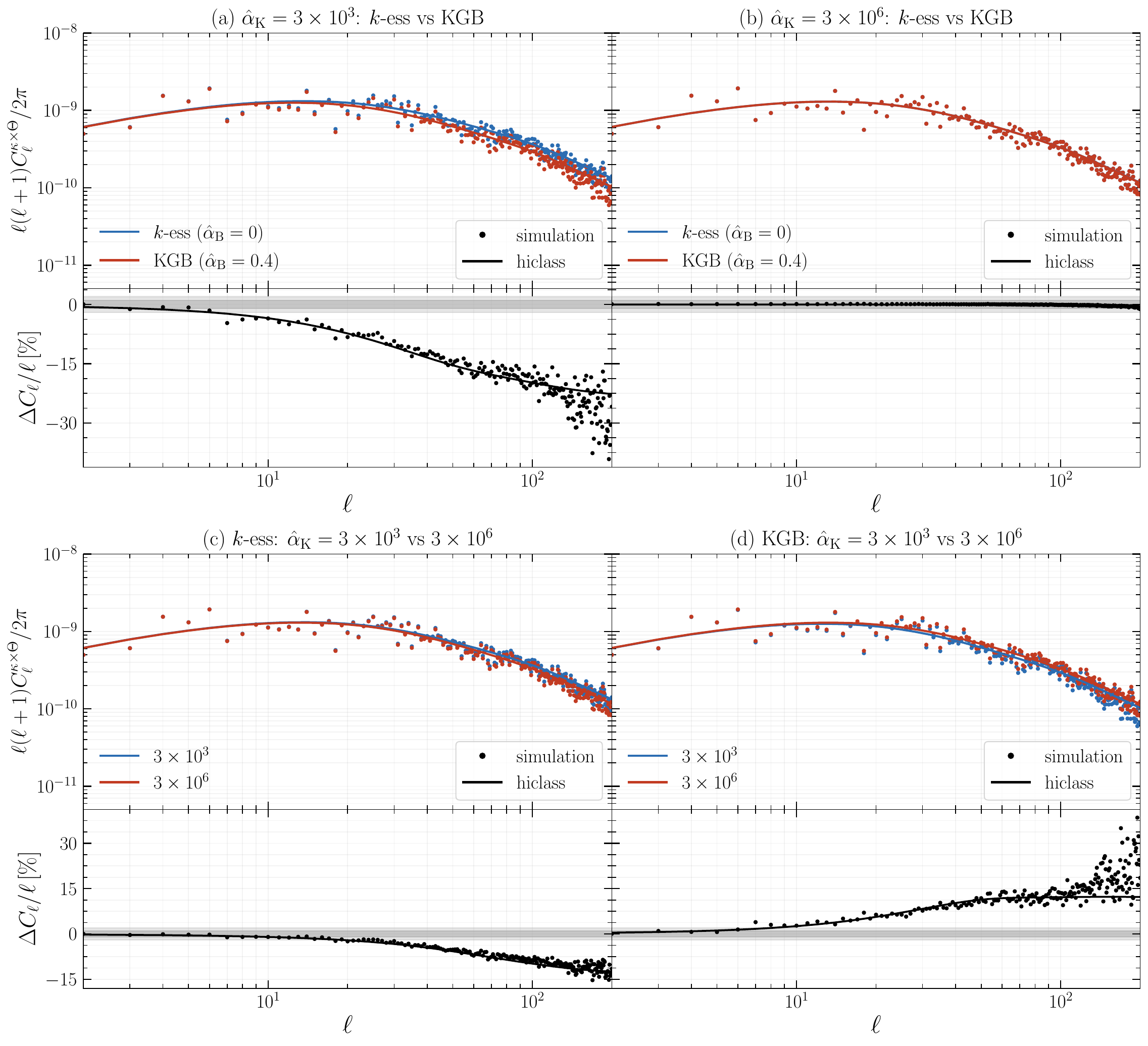}
    \caption{Angular cross-power spectrum between the weak lensing convergence and the ISW--RS signal at source redshift $z=0.86$, obtained from the simulations and from linear theory computed with \hiclass. The panel layout follows Fig.~\ref{fig:Cl_kappa}: panels (a,b) compare the $k$-essence ($\hat\alpha_{\rm B}=0$) and KGB ($\hat\alpha_{\rm B}=0.4$) models for $\hat\alpha_{\rm K}=3\times10^{3}$ and $\hat\alpha_{\rm K}=3\times10^{6}$, respectively, while panels (c,d) illustrate the dependence on $\hat\alpha_{\rm K}$ within each model separately. The lower panels show the fractional difference with respect to the chosen reference model.}
    \label{fig:Cl_kappaXisw}
\end{figure}

The resulting cross-power spectrum is shown in Fig.~\ref{fig:Cl_kappaXisw}. At large angular scales, the correlation is strong, as both signals are sourced by the same long-wavelength modes of the potential. Towards smaller scales,
nonlinear effects become important and differences between the models begin to
emerge. In particular, the KGB model exhibits a lower amplitude compared to $k$-essence, consistent with the suppression of the time derivative of the Weyl potential discussed above. At higher multipoles, nonlinear effects associated with the RS regime are expected to become important. However, the results of the simulation become increasingly noisy at high $\ell$, and we therefore restrict the analysis to $\ell \lesssim 200$, as shown in Fig.~\ref{fig:Cl_kappaXisw}. Within this range, no significant deviation between the linear predictions and the simulation results is observed.

\subsection*{Pencil-beam light cone}

In addition to the full-sky light cone extending to $\chi_s = 2015\,\mathrm{Mpc}/h$ $(z = 0.86)$,
we consider a pencil-beam light cone reaching $\chi_s \simeq 4690\,\mathrm{Mpc}/h$ $(z = 3.6)$.
In this case, the smaller sky coverage is traded for a larger radial depth, allowing us
to probe the same observables at higher redshift. To account for the incomplete sky coverage, we estimate the angular power spectra using the pseudo-$C_\ell$ formalism implemented in NaMaster \cite{Alonso:2018jzx}, where the observed spectra are corrected for mask-induced mode coupling via the mode coupling matrix $M_{\ell\ell'}$.

\begin{figure}[h]
    \centering
    \includegraphics[width=\linewidth]{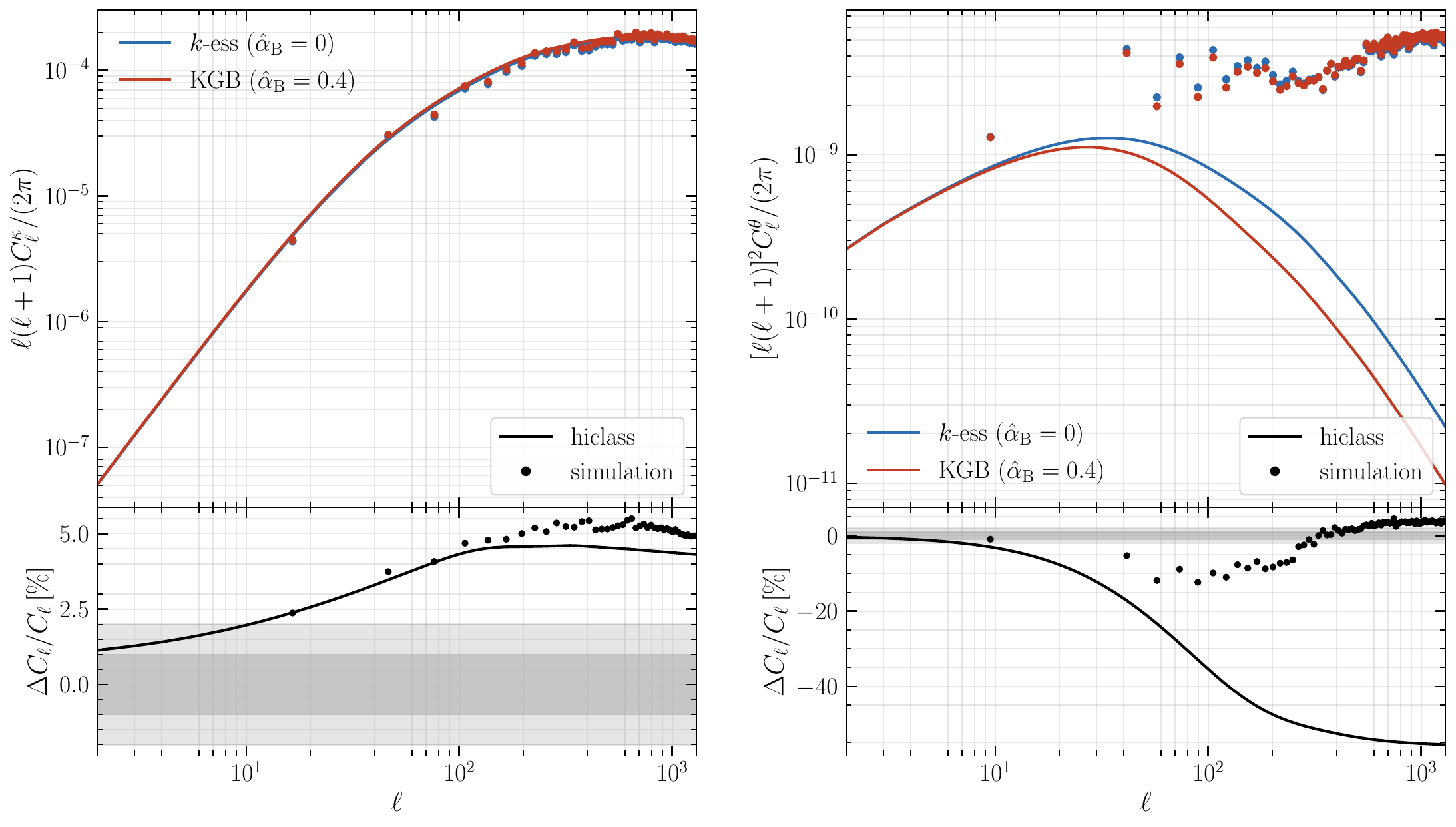}
    \caption{Angular power spectra of the convergence (\textit{left}) and ISW--RS signal (\textit{right}) at source redshift $z=3.6$, obtained from the pencil-beam light cone for $\hat\alpha_{\rm K}=3\times10^{3}$. The $k$-essence ($\hat\alpha_{\rm B}=0$) and KGB ($\hat\alpha_{\rm B}=0.4$) models are shown, including both the simulation results and the corresponding linear theory predictions computed with \hiclass. The lower panels show the fractional difference with respect to the $k$-essence model. The grey bands correspond to $1\%$ and $2\%$ fractional differences.}
    \label{fig:pencilBeam}
\end{figure}

Compared to the full-sky case, the pencil-beam light cone probes significantly
larger comoving distances, corresponding to higher redshifts. This affects the
amplitude of the observables. In the following, we focus on the weak lensing convergence and the ISW--RS signal, for which the differences between the models are more pronounced, while for the remaining observables the deviations are comparatively small and are therefore not shown. As shown in the left panel of Fig.~\ref{fig:pencilBeam}, the extended line-of-sight integration enhances the convergence signal by accumulating lensing contributions from a larger number of intervening structures, resulting in a higher overall amplitude than in the full-sky case. A similar enhancement is observed for the ISW--RS signal, as shown in the right panel of Fig.~\ref{fig:pencilBeam}. Although the time variation of the gravitational potentials is weaker at higher redshift, the larger radial
extent of the light cone increases the total integrated contribution. As a result, the ISW--RS amplitude is also enhanced in the pencil-beam case. At the same time, the differences between the KGB and $k$-essence models are
reduced compared to the full-sky results. This is particularly visible for the convergence, where the fractional deviation between the models is smaller. The reason is that the effect of dark energy clustering develops primarily at late times. By extending the light cone to higher redshift, a larger fraction of the signal originates from epochs where both models behave more similarly, which dilutes their relative differences in the projected spectra.

Finally, the increased contribution from larger comoving distances shifts the projection toward a more linear regime: at fixed multipole, larger $\chi$ corresponds approximately to smaller physical wavenumbers, $k\simeq \ell/\chi$, while structure formation is also less advanced at higher redshift. The model-dependent nonlinear corrections are therefore less pronounced, reducing the fractional KGB--$k$-essence difference generated by nonlinear evolution relative to the full-sky case. Nevertheless, the simulation results depart from the linear \hiclass predictions at sufficiently high multipoles because the projected signals still receive contributions from nonlinear structures, particularly at lower redshifts. This discrepancy is especially pronounced for the ISW--RS signal, since the simulations capture the nonlinear Rees--Sciama contribution arising from the nonlinear evolution of the gravitational potentials, which is absent from the linear 
\hiclass calculation.

\section{Conclusion}
\label{sec:conclusions}
In this work, we have investigated the predictions of the kinetic gravity braiding model for relativistic cosmological observables using light-cone outputs from \kgb simulations. We have focused on probes that depend directly on the gravitational field, including weak gravitational lensing, the ISW--RS effect, Shapiro time delay, and gravitational redshift, and compared the results to the corresponding $k$-essence limit and to the linear theory using \hiclass.

We have shown that braiding modifies both the amplitude and the time evolution of the gravitational potentials. The enhanced clustering of dark energy in KGB provides an additional source for the Weyl potential, increasing its amplitude and slowing its decay. For the representative model with $(\hat{\alpha}_{\rm K},\hat{\alpha}_{\rm B})=(3\times10^{3},0.4)$, the convergence spectrum is enhanced relative to $k$-essence by up to approximately $12\%$ at $\ell\sim10^{2}\text{--}10^{3}$, with only modest nonlinear departures from the \hiclass prediction at high multipoles. The ISW--RS signal exhibits a stronger scale dependence. At $\ell\lesssim30$, the simulations agree well with linear theory, whereas over the ISW-dominated intermediate range the slower decay of the Weyl potential suppresses the KGB signal relative to $k$-essence, with the difference reaching approximately $-30\%$. At higher multipoles, the transition from decaying to growing potentials occurs on larger scales in KGB than in $k$-essence, enhancing the nonlinear RS contribution. The simulated KGB signal consequently overtakes the $k$-essence result at $\ell\sim200\text{--}300$ and reaches an excess of order $10\%$ by $\ell\sim10^{3}$, while the corresponding linear KGB--$k$-essence difference predicted by \hiclass approaches a suppression of approximately $50\%$. Shapiro time delay and gravitational redshift show weaker responses at the few-percent level, with the Shapiro difference reaching approximately $7\%$.

We also find that kineticity affects dark energy clustering differently in KGB and $k$-essence. In the $k$-essence limit, the relevant transition is set by the sound horizon: increasing $\hat{\alpha}_{\rm K}$ lowers the scalar sound speed and shifts this transition to higher $k$, allowing dark energy to cluster over a broader range of wavenumbers. Braiding introduces the additional scale $k_{\rm B}$, below which the dynamics remains close to the $k$-essence limit and above which braiding modifies the evolution of the perturbations. At fixed $\hat{\alpha}_{\rm B}$, increasing $\hat{\alpha}_{\rm K}$ shifts both characteristic scales to higher $k$, extending the agreement with $k$-essence. Within KGB, however, it also strengthens the cancellation between the dominant contributions to the scalar-field density contrast, suppressing intermediate-scale clustering and reversing the corresponding $k$-essence trend, although in linear theory the dependence on $\hat{\alpha}_{\rm K}$ weakens again at high $k$. This scale-dependent response demonstrates that, unlike in $k$-essence, KGB clustering is not governed by the sound-horizon scale alone.

These results provide a basis for future comparisons with observations using tomographic predictions and realistic source-redshift distributions. Combining weak lensing and ISW-sensitive measurements could provide complementary tests of the scale-dependent signatures of kinetic gravity braiding in upcoming Stage-IV cosmological surveys.

\section*{Acknowledgements}
We would like to thank Camille Bonvin for helpful discussions and constructive feedback.
This work was completed in part at EuroHack24 with CSCS, part of the Open Hackathons programme. The authors would like to acknowledge OpenACC-Standard.org for their support. This work
was supported by a grant from the Swiss National Supercomputing
Centre (CSCS) under project ID sm97. Ahmad Nouri-Zonoz, Julian Adamek and Martin Kunz acknowledge
funding from the Swiss National Science Foundation. Julian Adamek acknowledges additional support from the Dr.\ Tomalla Foundation for Gravity Research. Some of the simulations in this work were performed on resources provided by Sigma2---the National
Infrastructure for High-Performance Computing and Data Storage in Norway. Farbod Hassani acknowledges funding from the Research Council of Norway (Young Talent Grant, Project No. 345334). Disclaimer: Co-funded by the European Union. Views and opinions expressed are, however, those of the author(s) only and do not necessarily reflect those of the European Union or the European Research Executive Agency. Neither the European Union nor the granting authority can be held responsible for them.

\appendix
\section{Computation of relativistic observables in the code}
\label{app:lc}
In this appendix, we describe the construction of the light-cone outputs and the post-processing procedure used to compute the observables. For further documentation and practical information, we refer the reader to the \gev user manual\footnote{\url{https://github.com/gevolution-code/gevolution-1.3}}.

In an observational context, a light cone corresponds to a null hypersurface representing all spacetime events that are visible to an observer at a given time. In the simulation, this hypersurface is specified through four geometric inputs: the comoving position of the observer, the observation time, a radial distance interval defining the look-back range, and a solid angle describing the survey geometry, which may correspond either to the full-sky or to a restricted field of view. In the latter case, the survey region is further characterised by an opening half-angle (in degrees) together with a direction vector, which can be provided in either Cartesian or polar coordinates. The configuration options controlling this construction are summarised in Table~\ref{tab:LCparams}.

The light-cone outputs are generated on the fly as the simulation evolves in time. All light-cone data are written in binary format: field quantities are stored in a custom raw binary \verb|.map| format, while particle light-cone outputs use the standard Gadget-2 binary format. In the following, we restrict the discussion to the light cones of the metric fields, as these form the basis for the computation of the relativistic observables considered in this work.

\begin{table}[h]
\centering
\renewcommand{\arraystretch}{1.25}
\begin{tabular}{|l|p{8cm}|}
\hline
\rowcolor{gray!20}
\textbf{Option} & \textbf{Description} \\
\hline

\verb|lightcone outputs = phi, Gadget2| &
Light-cone datasets to be written \\ \hline

\verb|lightcone vertex = 0,0,0| &
Observer position in comoving coordinates \\ \hline

\verb|lightcone redshift = 0| &
Redshift of the observation time \\ \hline

\verb|lightcone distance = 20, 500| &
Comoving look-back distance range in Mpc$/h$ \\ \hline

\verb|lightcone opening half-angle = 20| &
Maximum angle from the survey polar axis \\ \hline

\verb|lightcone direction = 0,0,1| &
Direction of the survey polar axis \\ \hline

\verb|lightcone Nside = 4, 1024| &
HEALPix angular-resolution range \\ \hline

\verb|lightcone pixel factor = 0.5| &
Adaptive angular sampling relative to the grid resolution \\ \hline

\verb|lightcone covering = 1| &
Overlap between consecutive metric-shell outputs \\ \hline

\verb|lightcone shell factor = 2| &
Radial shell sampling rate \\ \hline

\end{tabular}
\caption{Light-cone configuration parameters. The numerical values shown here are illustrative and do not correspond to the example discussed below.}
\label{tab:LCparams}
\end{table}

In the following, we describe the light-cone construction and analysis procedure step by step, and we illustrate each stage with an example taken from a simulation output. In this run we place the observer at the origin,
\verb|lightcone vertex = 0,0,0|, and consider a pencil-beam geometry with opening half-angle $30^\circ$ around the direction specified via \verb|lightcone direction = 1,1,1|.
We output metric light-cone shells out to a maximum comoving distance of $1000~\mathrm{Mpc}/h$ (\verb|lightcone distance = 1000|).
The simulation volume has a box size of $L=4032~\mathrm{Mpc}/h$ and uses $N_{\rm grid}=1200$ grid points per dimension.

\subsection{Criterion for writing light-cone shells}
\label{subsec:timeCrit}
The first practical question is when, during the evolution from the initial redshift, the code actually starts writing light-cone data to disk. In the code, this is decided locally at each cycle by comparing the user-requested light-cone distance range to a cycle-dependent radial interval bracketing the observer’s past light-cone radius at that simulation time. For this timing decision, the relevant inputs are the observation redshift (\verb|lightcone redshift|), which fixes the observation time, the requested look-back distance interval (\verb|lightcone distance|), and the covering factor (\verb|lightcone covering|), which sets the finite time-thickness of the recorded light cone and therefore controls when the requested distance range is first intersected.

In the code, all radii are handled in units of the box size $L$. The observation redshift (\verb|lightcone redshift|) fixes the conformal time of the observer, denoted $\tau_{\rm obs}$, while the simulation evolves at conformal times $\tau_i$ (cycle $i$). The corresponding infinitesimally thin past light-cone radius is therefore
\begin{equation}
r_{\star}(i)= \tau_{\rm obs}-\tau_i\, .
\end{equation}
Rather than recording an infinitely thin sphere at $r_{\star}$, the code constructs a finite \emph{radial recording window}
\begin{equation}
\label{eq:lc_window}
s_0(i)=(\tau_{\rm obs}-\tau_i)-\frac{1}{2}\,C\,\Delta\tau_i \, ,
\qquad
s_1(i)=(\tau_{\rm obs}-\tau_i)+\frac{1}{2}\,C\,\Delta\tau_{i-1}\, ,
\end{equation}

where $C$ is the covering factor and $\Delta\tau_i$ ($\Delta\tau_{i-1}$) are the current (previous) conformal-time step sizes. This window brackets the null surface over a finite time interval, enabling time interpolation and time derivatives in post-processing.

Let $[r_{\min},r_{\max}]$ be the requested radial range. The light-cone shells are written at cycle $i$ if and only if the cycle-dependent recording window overlaps the requested radial range, i.e.
\begin{equation}
\label{eq:lc_overlap}
[s_0(i),\,s_1(i)] \cap [r_{\min},\,r_{\max}] \neq \varnothing \, .
\end{equation}

\paragraph{Example case}
As we mentioned before, in the illustrative run considered here we use a cubic simulation volume of side length
$L=4032~\mathrm{Mpc}/h$ and request a maximum look-back distance of
\verb|lightcone distance = 1000|~$\mathrm{Mpc}/h$ (with $r_{\min}=0$). The code reports conformal time and radial distances in units of the box length, so the requested radial range can be written as
\begin{equation}
\frac{r_{\max}}{L}=\frac{1000}{4032}\simeq 0.248016 \, , \qquad
[r_{\min},r_{\max}] \;\rightarrow\; \left[0,0.248016\right] \, .
\end{equation}
Moreover, we used \verb|lightcone covering = 2|, implying that the metric recording window in Eq.~\eqref{eq:lc_window} spans two neighbouring time steps around the null hypersurface. Light-cone output begins at the first cycle for which the overlap condition in Eq.~\eqref{eq:lc_overlap} is satisfied. In our case, this occurs at cycle $i=108$, where $s_0(108)=0.226863$ and $s_1(108)=0.295191$, such that
\begin{equation}
[s_0(108),\,s_1(108)] \cap \left[0,\,\frac{r_{\max}}{L}\right] \neq \varnothing\, .
\end{equation}
The code therefore starts writing light-cone shells from that cycle onward. Note that at this first output cycle the central radius $r_{\star}(108)=\tau_{\rm obs}-\tau_{108}$ can still exceed $r_{\max}/L$; the output is triggered by the finite-thickness window $[s_0,s_1]$ intersecting the requested range, not by $r_{\star}$ alone.

\subsection{HEALPix resampling of lattice fields on spherical shells}
Once the timing criterion in Section~\ref{subsec:timeCrit} is satisfied at a given cycle, the code constructs the light-cone output as a set of \emph{discrete} spherical shells centred on the observer. The finite interval $[s_0(i),s_1(i)]$ introduced above should be understood as a \emph{selection window} in radius: the code outputs a collection of two-dimensional spherical surfaces with radii $r_{\rm shell}$ that fall within the window. The radial spacing between consecutive shells is fixed and given by
\begin{equation}
\Delta r = \frac{1}{N_{\rm grid}\,f_{\rm shell}}\, ,
\end{equation}
where $N_{\rm grid}$ is the number of lattice points per dimension and $f_{\rm shell}$ is the shell factor (\verb|lightcone shell factor|) set by the user.

In the next step, for each shell radius $r_{\rm shell}$, the two-dimensional sphere is pixelised using the HEALPix framework. In the full-sky case the number of pixels is $N_{\rm pix}=12\,N_{\rm side}^2$. The angular resolution of this pixelisation is controlled through the option \verb|lightcone Nside|, for which the user provides a minimum and maximum value. If a single value is specified, the output maps are produced at fixed angular resolution for all shells. Otherwise, the code employs an adaptive angular resolution based on \verb|lightcone pixel factor| which specifies a spatial sampling on the shell relative to the Cartesian grid.

After the code has chosen the HEALPix resolution $N_{\rm side}$ for a given shell (either fixed or adaptively via the pixel factor), it proceeds by evaluating the metric field at the centres of the HEALPix pixels. Each pixel centre defines a line of sight $\hat{\mathbf n}_p$; at shell radius $r_{\rm shell}$ this direction corresponds to the comoving sampling position
\begin{equation}
\mathbf x_p(r_{\rm shell}) = \mathbf x_{\rm obs} + r_{\rm shell}\,\hat{\mathbf n}_p\, ,
\end{equation}
with an additional rotation applied if a pointing direction is specified. Periodic boundary conditions are enforced by wrapping $\mathbf x_p$ back into the simulation volume.

The metric fields are represented on the Cartesian simulation lattice; hence, $\mathbf x_p$ will in general not coincide with a lattice site. The field value assigned to pixel $p$ is obtained by interpolating the lattice field to $\mathbf x_p$. This is done by identifying the unique grid cell that contains $\mathbf x_p$ and performing cloud-in-cell (CIC) interpolation from the field values at the eight vertices of that cell. Repeating this procedure for all pixels $p=1,\dots,N_{\rm pix}$ yields the two-dimensional HEALPix map for that shell radius. The complete light-cone output for a given cycle consists of the set of such maps for all radii selected in that cycle. Accordingly, each light-cone output file \verb|lightcone_XXXX.map| (one file per simulation cycle) stores the collection of HEALPix maps corresponding to all discrete shell radii $r_{\rm shell}$ selected in that cycle, i.e. the set of two-dimensional pixelised spherical shells whose radii fall within the cycle-dependent recording window $[s_0(i),s_1(i)]$. Note that when adaptive angular resolution is used, the HEALPix resolution (and thus the pixelisation) may differ from shell to shell.

This procedure is repeated at every simulation cycle, whenever the overlap criterion in Eq.~\eqref{eq:lc_overlap} is satisfied: the code selects the shell radii within the corresponding window $[s_0(i),s_1(i)]$, resamples the lattice fields onto the associated HEALPix pixelisations, and writes the resulting collection of shell maps to disk as a file \verb|lightcone_XXXX.map| for that cycle. In addition, the code writes an accompanying information file for each light cone, which records the geometric parameters and, for every cycle that satisfies the timing criterion, the corresponding radial recording window. In this file the bounds $s_0(i)$ and $s_1(i)$ are reported as \texttt{metric\_inner} and \texttt{metric\_outer}, respectively, providing a per-cycle summary of the radial interval for which metric shells have been output.

\begin{figure}[h]
    \centering
    \includegraphics[width=\linewidth]{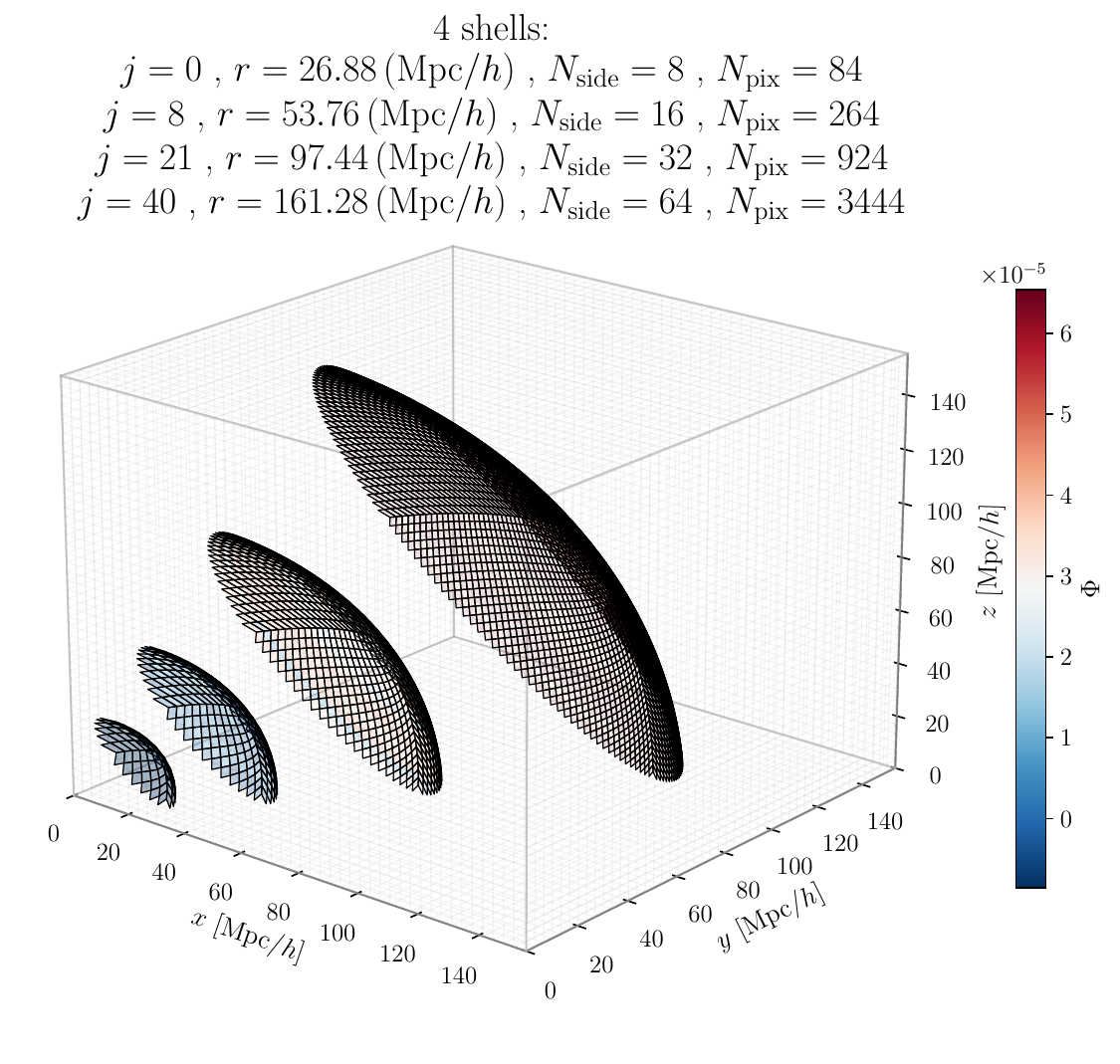}
\caption{Selected light-cone shells for the metric potential $\Phi$ near the observer, taken from the last simulation cycle (which corresponds to the final light-cone output written) and indexed by $j$. The change in patch size indicates the adaptive HEALPix resolution $N_{\rm side}$ with radius.}
    \label{fig:shells}
\end{figure}

\paragraph{Example case.}
For visualisation, Fig.~\ref{fig:shells} shows a subset of the pixelised spherical shells recorded at the final simulation step, here cycle 117 (\verb|lightcone_0117.map|).
Each coloured patch corresponds to one HEALPix pixel on a given shell, while the shells are centred on the observer at the light-cone vertex. In the figure, $j$ labels the shell index within that output cycle, i.e.\ $j$ enumerates the discrete radii $r_{\rm shell}$ that fall into the recording window $[s_0(i),s_1(i)]$. 
As can be seen, the angular resolution varies from shell to shell: shells at different radii are written with different HEALPix resolutions $N_{\rm side}$, reflecting the adaptive angular sampling controlled by \verb|lightcone pixel factor|. For clarity, we do not plot every shell contained in the cycle; instead, we select a representative set of shells, with the spacing in $j$ chosen to highlight radii at which the code changes $N_{\rm side}$. In particular, the number of pixels increases with $N_{\rm side}$, so that shells written at larger $N_{\rm side}$ appear more finely pixelised than those at smaller $N_{\rm side}$.

\subsection{Post-processing and map construction with \texttt{lcmap}}
The light-cone files \verb|lightcone_XXXX.map| produced by the code contain the metric fields sampled on discrete HEALPix shells, but they are not yet fully in a form that is directly usable for line-of-sight integrals for calculating different observables. The companion tool \texttt{lcmap} performs this post-processing: it reads the metric light-cone output and constructs the \emph{local} endpoint potential terms (e.g.\ the gravitational redshift contribution $\Psi_o-\Psi_s$) as well as the \emph{integrated} potential contributions along the line of sight (e.g.\ the Shapiro time delay and the ISW--RS term) entering Eq.~\eqref{eq:3Dposition}.

\texttt{lcmap} is executed with the simulation settings file together with either a source distance or source redshift (and optionally, one or more light-cone IDs):
\begin{equation}
\texttt{lcmap -s <settingsfile> \;\Big( -d <}\chi_s\texttt{,\dots> \;\;|\;\; -z <}z_s\texttt{,\dots>\Big)\; -l <ID1,ID2,\dots>}\, .
\end{equation}
The output is written as HEALPix FITS files following the naming conventions specified in the simulation settings.

To evaluate the contributions entering Eq.~\eqref{eq:3Dposition}, \texttt{lcmap} traverses the light cone radially outward and maintains a set of accumulator maps on the sky, one for each local or integrated contribution of interest. Denoting a generic accumulated quantity by $A(\hat{\mathbf n})$ (e.g.\ the Shapiro time delay, the ISW--RS term, or lensing-related line-of-sight integrals), the code updates the accumulator shell by shell according to the schematic rule
\begin{equation}
\label{eq:lcmap_update_generic}
A(\hat{\mathbf n}) \;\leftarrow\; A(\hat{\mathbf n}) \;+\; \mathcal{I}_A(r_j,\hat{\mathbf n})\,\Delta r_j\, ,
\end{equation}
where $\mathcal{I}_A$ denotes the corresponding integrand evaluated on the shell at radius $r_j$, and $\Delta r_j$ is the radial step between successive shells. In practice, the required shell values are evaluated at the appropriate light-cone time by interpolating between two neighbouring light-cone cycles, but this does not change the schematic structure of the radial accumulation.

For a requested source distance $\chi_s$ (or redshift), \texttt{lcmap} advances the accumulation until the shell radii bracket the target distance and then writes the corresponding output maps.\footnote{In general, the requested $\chi_s$ (or $z_s$) does not coincide exactly with one of the discrete shell radii, so the output at the target distance is obtained by interpolation between the neighbouring shells that bracket $\chi_s$.}

As we showed in the previous section, when adaptive angular resolution is enabled, the shell maps may be stored at different HEALPix resolutions $N_{\rm side}(r_j)$. During the radial traversal of the light cone, the code keeps track of the highest HEALPix resolution encountered and resizes the accumulator arrays whenever a shell with larger $N_{\rm side }$
 appears. However, previously accumulated contributions are not immediately interpolated to the finer grid. Instead, the accumulators retain the values computed at their native resolutions during the integration. Once the radial traversal reaches the requested source distance and the final output maps are constructed, the code performs a HEALPix interpolation that transfers the accumulated contributions from all intermediate resolutions to the finest grid encountered along the light cone. The final maps are therefore written at this highest angular resolution.

In Fig.~\ref{fig:shells}, we illustrate how the angular resolution of the light-cone shells is increased with distance from the observer. As the shells become larger, a higher HEALPix resolution is used to maintain an adequate sampling of the lattice fields across the observed patch.

\section{Convergence tests}
\label{app:conv_test}
\begin{figure}[H]
  \centering
  \includegraphics[width=0.93\textwidth]{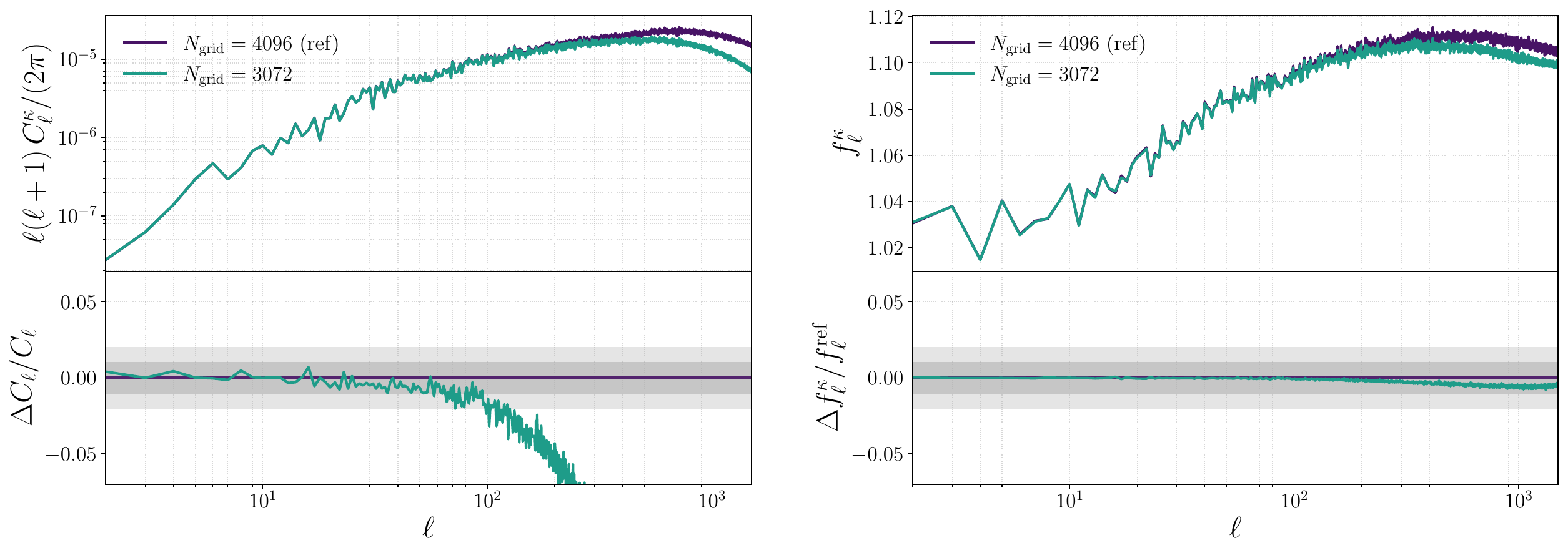}\par\vspace{0.0em}
  \includegraphics[width=0.93\textwidth]{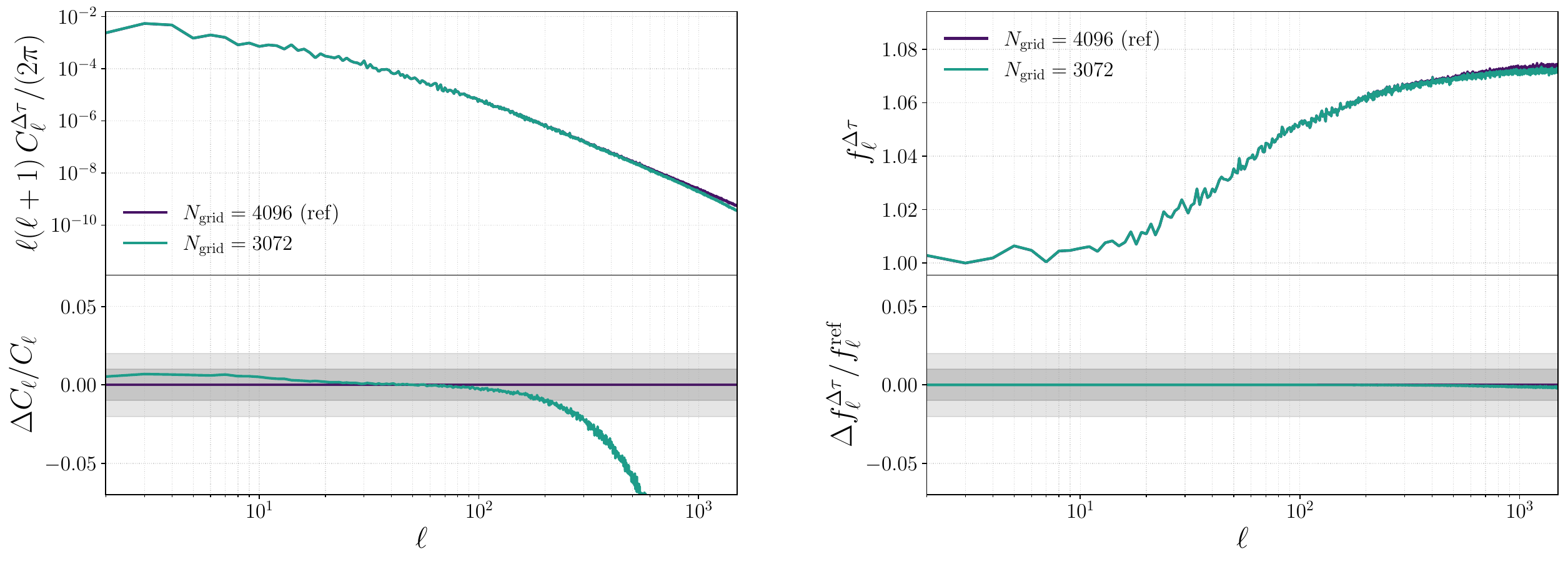}\par\vspace{0.0em}
  \includegraphics[width=0.93\textwidth]{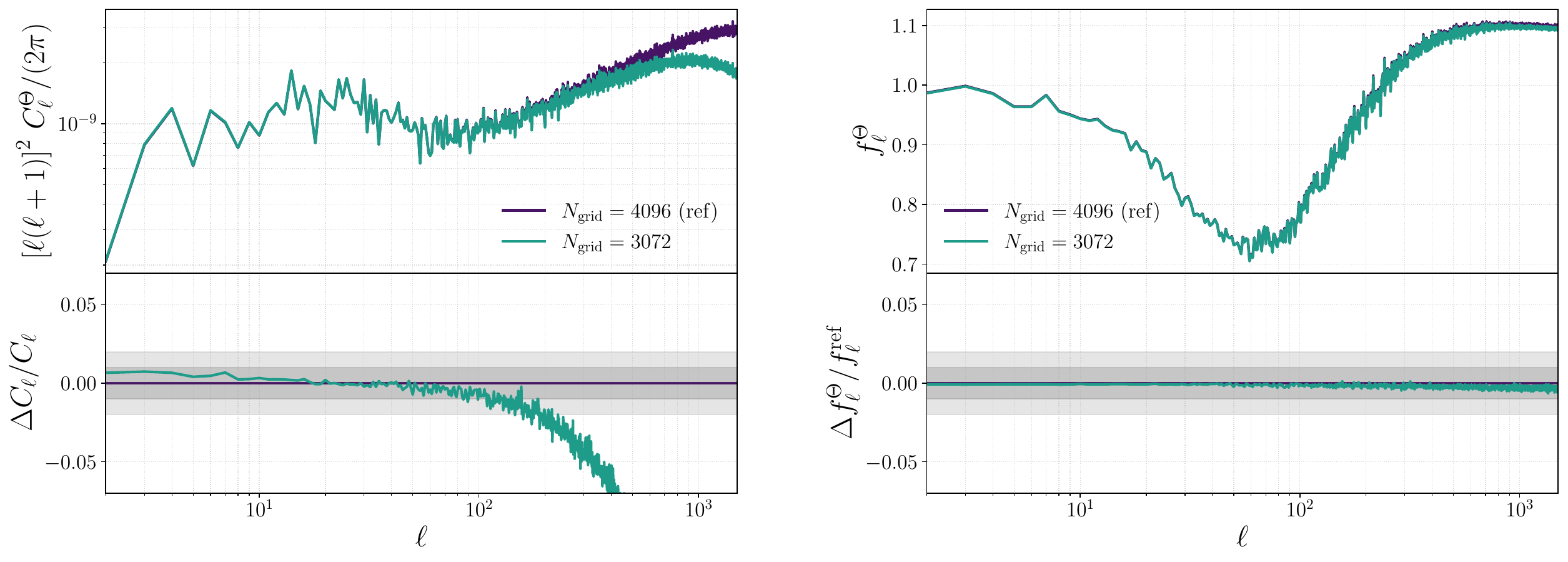}\par\vspace{0.0em}
  \includegraphics[width=0.93\textwidth]{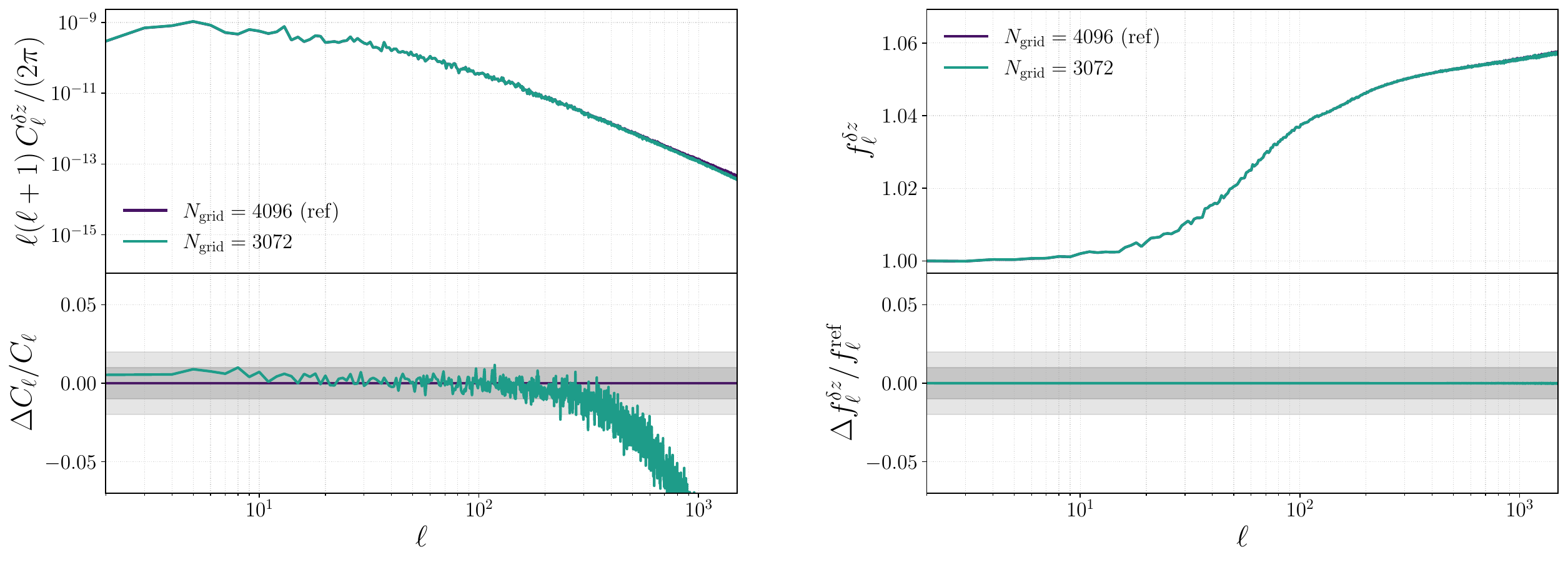}
\caption{Convergence tests of the full-sky light-cone angular power spectra at $z=0.86$ in a box of size $L_{\rm box}=4032~{\rm Mpc}/h$. \textit{Left} panels: KGB auto-spectra $C_\ell$ for $\hat\alpha_{\rm K}=3 \times 10^3$ and $\hat\alpha_{\rm B}=0.4$, with lower subpanels showing $\Delta C_\ell/C_\ell$ relative to the $N_{\rm grid}=4096$ reference. \textit{Right} panels: convergence of the model ratio $f_\ell \equiv C_\ell^{\rm KGB}/C_\ell^{k\rm ess}$, with lower subpanels showing $\Delta f_\ell/f_\ell^{\rm ref}$ relative to the same reference. }
  \label{fig:convergence}
\end{figure}

In Fig.~\ref{fig:convergence}, we present a numerical convergence test for our full-sky light-cone angular power spectra at $z=0.86$ in a fixed comoving box of size $L_{\rm box}=4032~{\rm Mpc}/h$. We compare the fiducial simulation, with $N_{\rm grid}=3072$ mesh points per dimension and $N_{\rm pcl}=3072^3$ particles, against a higher-resolution reference simulation with $N_{\rm grid}=4096$ and $N_{\rm pcl}=4096^3$. The left panels show the auto-spectra $C_\ell$ for the KGB model ($\hat\alpha_{\rm K}=3\times10^3$, $\hat\alpha_{\rm B}=0.4$), with the subpanels reporting the relative deviation $\Delta C_\ell/C_\ell$ with respect to the $N_{\rm grid}=4096$ reference. The right panels show the convergence of the ratio between the KGB and $k$-essence angular power spectra, $f_\ell \equiv C_\ell^{\rm KGB}/C_\ell^{k\text{ess}}$, for $N_{\rm grid}=3072$ and $N_{\rm grid}=4096$, with the subpanels showing the relative deviation $\Delta f_\ell/f_\ell^{\rm ref}$ with respect to the $N_{\rm grid}=4096$ reference. 

In Table~\ref{tab:convergence_limits}, we summarise the corresponding values of $\ell_{\max}$ up to which the auto-spectra and the fractional differences between the two models satisfy the conservative $2\%$ criterion. Although the impact of resolution is evident in the absolute spectra $C_\ell$, it largely cancels in the ratio $f_\ell$, indicating that the fractional differences between the two models can be trusted over a significantly wider range of multipoles than the individual auto-spectra.
\begin{table}[H]
\centering
\caption{Convergence limits for our fiducial simulations with
$N_{\rm grid}=3072$, evaluated relative to the higher-resolution
reference simulation with $N_{\rm grid}=4096$.
For each observable we report the maximum multipole $\ell_{\max}$ up to which (i) the KGB auto-spectrum $C_\ell$ agrees with the reference at the $2\%$ level, and (ii) the KGB-to-$k$-essence ratio $f_\ell$ is converged to within $2\%$ relative to the $N_{\rm grid} = 4096$ reference.
}
\vspace{0.5em}
\begin{tabular}{lcc}
\hline\hline
Observable & $\ell_{\max}$ for $C_\ell$ (KGB) & $\ell_{\max}$ for $f_\ell$ (KGB vs $k$-ess) \\
\hline
Convergence $\kappa$        & $\sim 100$  & $\sim 1000$ \\
Shapiro time delay               & $\sim 300$ & $\sim 1000$ \\
ISW--RS                        & $\sim 200$ & $\sim 1000$ \\
Gravitational redshift            & $\sim 400$ & $\sim 1000$ \\
\hline\hline
\end{tabular}
\label{tab:convergence_limits}
\end{table}

\newpage

 \bibliographystyle{JHEP}
\bibliography{biblio.bib}

@article{Linder:2002et,
    author = "Linder, Eric V.",
    title = "{Exploring the expansion history of the universe}",
    eprint = "astro-ph/0208512",
    archivePrefix = "arXiv",
    doi = "10.1103/PhysRevLett.90.091301",
    journal = "Phys. Rev. Lett.",
    volume = "90",
    pages = "091301",
    year = "2003"
}

@article{Chevallier:2000qy,
    author = "Chevallier, Michel and Polarski, David",
    title = "{Accelerating universes with scaling dark matter}",
    eprint = "gr-qc/0009008",
    archivePrefix = "arXiv",
    doi = "10.1142/S0218271801000822",
    journal = "Int. J. Mod. Phys. D",
    volume = "10",
    pages = "213--224",
    year = "2001"
}

@article{DESI:2025fii,
    author = "Lodha, K. and others",
    collaboration = "DESI",
    title = "{Extended Dark Energy analysis using DESI DR2 BAO measurements}",
    eprint = "2503.14743",
    archivePrefix = "arXiv",
    primaryClass = "astro-ph.CO",
    reportNumber = "FERMILAB-PUB-25-0164-PPD",
    month = "3",
    year = "2025"
}

@article{Bellini:2014fua,
    author = "Bellini, Emilio and Sawicki, Ignacy",
    title = "{Maximal freedom at minimum cost: linear large-scale structure in general modifications of gravity}",
    eprint = "1404.3713",
    archivePrefix = "arXiv",
    primaryClass = "astro-ph.CO",
    doi = "10.1088/1475-7516/2014/07/050",
    journal = "JCAP",
    volume = "07",
    pages = "050",
    year = "2014"
}

@article{Adamek:2016zes,
    author = "Adamek, Julian and Daverio, David and Durrer, Ruth and Kunz, Martin",
    title = "{gevolution: a cosmological N-body code based on General Relativity}",
    eprint = "1604.06065",
    archivePrefix = "arXiv",
    primaryClass = "astro-ph.CO",
    doi = "10.1088/1475-7516/2016/07/053",
    journal = "JCAP",
    volume = "07",
    pages = "053",
    year = "2016"
}

@article{Hassani:2019lmy,
    author = "Hassani, Farbod and Adamek, Julian and Kunz, Martin and Vernizzi, Filippo",
    title = "{$k$-evolution: a relativistic N-body code for clustering dark energy}",
    eprint = "1910.01104",
    archivePrefix = "arXiv",
    primaryClass = "astro-ph.CO",
    doi = "10.1088/1475-7516/2019/12/011",
    journal = "JCAP",
    volume = "12",
    pages = "011",
    year = "2019"
}

@article{Hassani:2019wed,
    author = "Hassani, Farbod and L'Huillier, Benjamin and Shafieloo, Arman and Kunz, Martin and Adamek, Julian",
    title = "{Parametrising non-linear dark energy perturbations}",
    eprint = "1910.01105",
    archivePrefix = "arXiv",
    primaryClass = "astro-ph.CO",
    doi = "10.1088/1475-7516/2020/04/039",
    journal = "JCAP",
    volume = "04",
    pages = "039",
    year = "2020"
}

@article{Adamek:2015eda,
    author = "Adamek, Julian and Daverio, David and Durrer, Ruth and Kunz, Martin",
    title = "{General relativity and cosmic structure formation}",
    eprint = "1509.01699",
    archivePrefix = "arXiv",
    primaryClass = "astro-ph.CO",
    doi = "10.1038/nphys3673",
    journal = "Nature Phys.",
    volume = "12",
    pages = "346--349",
    year = "2016"
}

@article{Horndeski:1974wa,
    author = "Horndeski, Gregory Walter",
    title = "{Second-order scalar-tensor field equations in a four-dimensional space}",
    doi = "10.1007/BF01807638",
    journal = "Int. J. Theor. Phys.",
    volume = "10",
    pages = "363--384",
    year = "1974"
}

@article{Zumalacarregui:2016pph,
    author = "Zumalac\'arregui, Miguel and Bellini, Emilio and Sawicki, Ignacy and Lesgourgues, Julien and Ferreira, Pedro G.",
    title = "{hi\_class: Horndeski in the Cosmic Linear Anisotropy Solving System}",
    eprint = "1605.06102",
    archivePrefix = "arXiv",
    primaryClass = "astro-ph.CO",
    reportNumber = "NORDITA-2016-41",
    doi = "10.1088/1475-7516/2017/08/019",
    journal = "JCAP",
    volume = "08",
    pages = "019",
    year = "2017"
}

@article{Bellini:2019syt,
    author = "Bellini, Emilio and Sawicki, Ignacy and Zumalac{\'a}rregui, Miguel",
    title = "{hi{\_}class: Background Evolution, Initial Conditions and Approximation Schemes}",
    eprint = "1909.01828",
    archivePrefix = "arXiv",
    primaryClass = "astro-ph.CO",
    doi = "10.1088/1475-7516/2020/02/008",
    journal = "JCAP",
    volume = "02",
    pages = "008",
    year = "2020"
}

@article{Deffayet:2010qz,
    author = "Deffayet, Cedric and Pujolas, Oriol and Sawicki, Ignacy and Vikman, Alexander",
    title = "{Imperfect Dark Energy from Kinetic Gravity Braiding}",
    eprint = "1008.0048",
    archivePrefix = "arXiv",
    primaryClass = "hep-th",
    reportNumber = "CERN-PH-TH-2010-166",
    doi = "10.1088/1475-7516/2010/10/026",
    journal = "JCAP",
    volume = "10",
    pages = "026",
    year = "2010"
}

@article{Hassani:2020buk,
    author = "Hassani, Farbod and Adamek, Julian and Kunz, Martin",
    title = "{Clustering dark energy imprints on cosmological observables of the gravitational field}",
    eprint = "2007.04968",
    archivePrefix = "arXiv",
    primaryClass = "astro-ph.CO",
    doi = "10.1093/mnras/staa3589",
    journal = "Mon. Not. Roy. Astron. Soc.",
    volume = "500",
    number = "4",
    pages = "4514--4529",
    year = "2020"
}

@article{PhysRevD.84.043516,
  title = {Linear power spectrum of observed source number counts},
  author = {Challinor, Anthony and Lewis, Antony},
  journal = {Phys. Rev. D},
  volume = {84},
  issue = {4},
  pages = {043516},
  numpages = {14},
  year = {2011},
  month = {Aug},
  publisher = {American Physical Society},
  doi = {10.1103/PhysRevD.84.043516},
  url = {https://link.aps.org/doi/10.1103/PhysRevD.84.043516}
}

@article{PhysRevD.80.083514,
  title = {New perspective on galaxy clustering as a cosmological probe: General relativistic effects},
  author = {Yoo, Jaiyul and Fitzpatrick, A. Liam and Zaldarriaga, Matias},
  journal = {Phys. Rev. D},
  volume = {80},
  issue = {8},
  pages = {083514},
  numpages = {10},
  year = {2009},
  month = {Oct},
  publisher = {American Physical Society},
  doi = {10.1103/PhysRevD.80.083514},
  url = {https://link.aps.org/doi/10.1103/PhysRevD.80.083514}
}

@article{10.1093/mnras/sty3206,
    author = {Breton, Michel-Andrès and Rasera, Yann and Taruya, Atsushi and Lacombe, Osmin and Saga, Shohei},
    title = {Imprints of relativistic effects on the asymmetry of the halo cross-correlation function: from linear to non-linear scales},
    journal = {Monthly Notices of the Royal Astronomical Society},
    volume = {483},
    number = {2},
    pages = {2671-2696},
    year = {2018},
    month = {11},
    issn = {0035-8711},
    doi = {10.1093/mnras/sty3206},
    url = {https://doi.org/10.1093/mnras/sty3206},
    eprint = {https://academic.oup.com/mnras/article-pdf/483/2/2671/27201436/sty3206.pdf},
}

@article{PhysRevD.101.023512,
  title = {Ray tracing the integrated Sachs-Wolfe effect through the light cones of the dark energy universe simulation-full universe runs},
  author = {Adamek, Julian and Rasera, Yann and Corasaniti, Pier Stefano and Alimi, Jean-Michel},
  journal = {Phys. Rev. D},
  volume = {101},
  issue = {2},
  pages = {023512},
  numpages = {13},
  year = {2020},
  month = {Jan},
  publisher = {American Physical Society},
  doi = {10.1103/PhysRevD.101.023512},
  url = {https://link.aps.org/doi/10.1103/PhysRevD.101.023512}
}

@article{PhysRevD.84.063505,
  title = {What galaxy surveys really measure},
  author = {Bonvin, Camille and Durrer, Ruth},
  journal = {Phys. Rev. D},
  volume = {84},
  issue = {6},
  pages = {063505},
  numpages = {16},
  year = {2011},
  month = {Sep},
  publisher = {American Physical Society},
  doi = {10.1103/PhysRevD.84.063505},
  url = {https://link.aps.org/doi/10.1103/PhysRevD.84.063505}
}

@article{Nouri-Zonoz:2025cul,
    author = "Nouri-Zonoz, Ahmad and Hassani, Farbod and Bellini, Emilio and Kunz, Martin",
    title = "{KGB-evolution: a relativistic $N$-body code for kinetic gravity braiding models}",
    eprint = "2511.04676",
    archivePrefix = "arXiv",
    primaryClass = "astro-ph.CO",
    month = "11",
    year = "2025"
}

@article{Kohlinger:2017sxk,
    author = {K{\"o}hlinger, F. and others},
    title = "{KiDS-450: The tomographic weak lensing power spectrum and constraints on cosmological parameters}",
    eprint = "1706.02892",
    archivePrefix = "arXiv",
    primaryClass = "astro-ph.CO",
    doi = "10.1093/mnras/stx1820",
    journal = "Mon. Not. Roy. Astron. Soc.",
    volume = "471",
    number = "4",
    pages = "4412--4435",
    year = "2017"
}

@article{Becker:2012qe,
    author = "Becker, Matthew R.",
    title = "{CALCLENS: Weak Lensing Simulations for Large-area Sky Surveys and Second-order Effects in Cosmic Shear Power Spectra}",
    eprint = "1210.3069",
    archivePrefix = "arXiv",
    primaryClass = "astro-ph.CO",
    month = "10",
    year = "2012"
}

@misc{EUCLID:2011zbd,
    author = "Laureijs, R. and others",
    collaboration = "EUCLID",
    title = "{Euclid Definition Study Report}",
    eprint = "1110.3193",
    archivePrefix = "arXiv",
    primaryClass = "astro-ph.CO",
    reportNumber = "ESA-SRE(2011)12",
    month = "10",
    year = "2011"
}

@misc{Akeson:2019biv,
    author = "Akeson, Rachel and others",
    title = "{The Wide Field Infrared Survey Telescope: 100 Hubbles for the 2020s}",
    eprint = "1902.05569",
    archivePrefix = "arXiv",
    primaryClass = "astro-ph.IM",
    month = "2",
    year = "2019"
}

@misc{LSSTScience:2009jmu,
    author = "Abell, Paul A. and others",
    collaboration = "LSST Science, LSST Project",
    title = "{LSST Science Book, Version 2.0}",
    eprint = "0912.0201",
    archivePrefix = "arXiv",
    primaryClass = "astro-ph.IM",
    reportNumber = "FERMILAB-TM-2495-A, SLAC-R-1031",
    month = "12",
    year = "2009"
}

@article{Hu:2013twa,
    author = "Hu, Bin and Raveri, Marco and Frusciante, Noemi and Silvestri, Alessandra",
    title = "{Effective Field Theory of Cosmic Acceleration: an implementation in CAMB}",
    eprint = "1312.5742",
    archivePrefix = "arXiv",
    primaryClass = "astro-ph.CO",
    doi = "10.1103/PhysRevD.89.103530",
    journal = "Phys. Rev. D",
    volume = "89",
    number = "10",
    pages = "103530",
    year = "2014"
}

@article{Armendariz-Picon:2000nqq,
    author = "Armendariz-Picon, C. and Mukhanov, Viatcheslav F. and Steinhardt, Paul J.",
    title = "{A Dynamical solution to the problem of a small cosmological constant and late time cosmic acceleration}",
    eprint = "astro-ph/0004134",
    archivePrefix = "arXiv",
    doi = "10.1103/PhysRevLett.85.4438",
    journal = "Phys. Rev. Lett.",
    volume = "85",
    pages = "4438--4441",
    year = "2000"
}

@article{Sawicki:2012re,
    author = "Sawicki, Ignacy and Saltas, Ippocratis D. and Amendola, Luca and Kunz, Martin",
    title = "{Consistent perturbations in an imperfect fluid}",
    eprint = "1208.4855",
    archivePrefix = "arXiv",
    primaryClass = "astro-ph.CO",
    doi = "10.1088/1475-7516/2013/01/004",
    journal = "JCAP",
    volume = "01",
    pages = "004",
    year = "2013"
}

@article{Lepori:2020ifz,
    author = "Lepori, Francesca and Adamek, Julian and Durrer, Ruth and Clarkson, Chris and Coates, Louis",
    title = "{Weak-lensing observables in relativistic N-body simulations}",
    eprint = "2002.04024",
    archivePrefix = "arXiv",
    primaryClass = "astro-ph.CO",
    doi = "10.1093/mnras/staa2024",
    journal = "Mon. Not. Roy. Astron. Soc.",
    volume = "497",
    number = "2",
    pages = "2078--2095",
    year = "2020"
}

@article{Magi:2026xcy,
    author = "Magi, Matteo and Lepori, Francesca and Adamek, Julian",
    title = "{Perturbative and numerical study of nonlinear relativistic effects in weak lensing}",
    eprint = "2603.24179",
    archivePrefix = "arXiv",
    primaryClass = "astro-ph.CO",
    month = "3",
    year = "2026"
}

@article{Bartelmann:1999yn,
    author = "Bartelmann, Matthias and Schneider, Peter",
    title = "{Weak gravitational lensing}",
    eprint = "astro-ph/9912508",
    archivePrefix = "arXiv",
    doi = "10.1016/S0370-1573(00)00082-X",
    journal = "Phys. Rept.",
    volume = "340",
    pages = "291--472",
    year = "2001"
}

@article{Alsing:2015zca,
    author = "Alsing, Justin and Heavens, Alan and Jaffe, Andrew H. and Kiessling, Alina and Wandelt, Benjamin and Hoffmann, Till",
    title = "{Hierarchical Cosmic Shear Power Spectrum Inference}",
    eprint = "1505.07840",
    archivePrefix = "arXiv",
    primaryClass = "astro-ph.CO",
    doi = "10.1093/mnras/stv2501",
    journal = "Mon. Not. Roy. Astron. Soc.",
    volume = "455",
    number = "4",
    pages = "4452--4466",
    year = "2016"
}

@article{Mandelbaum:2018,
  author = {Mandelbaum, Rachel},
  title = {Weak Gravitational Lensing for Precision Cosmology},
  journal = {Ann. Rev. Astron. Astrophys.},
  volume = {56},
  pages = {393-433},
  year = {2018},
  doi = {10.1146/annurev-astro-081817-051928},
  eprint = {1710.03235},
  archivePrefix = {arXiv},
  primaryClass = {astro-ph.CO}
}

@article{vanWaerbeke:2000rm,
    author = "van Waerbeke, Ludovic and others",
    title = "{Detection of correlated galaxy ellipticities on CFHT data: First evidence for gravitational lensing by large scale structures}",
    eprint = "astro-ph/0002500",
    archivePrefix = "arXiv",
    journal = "Astron. Astrophys.",
    volume = "358",
    pages = "30--44",
    year = "2000"
}

@article{Schmidt_2012,
doi = {10.1088/2041-8205/744/2/L22},
url = {https://doi.org/10.1088/2041-8205/744/2/L22},
year = {2011},
month = {dec},
publisher = {The American Astronomical Society},
volume = {744},
number = {2},
pages = {L22},
author = {Schmidt, Fabian and Leauthaud, Alexie and Massey, Richard and Rhodes, Jason and George, Matthew R. and Koekemoer, Anton M. and Finoguenov, Alexis and Tanaka, Masayuki},
title = {A DETECTION OF WEAK-LENSING MAGNIFICATION USING GALAXY SIZES AND MAGNITUDES},
journal = {The Astrophysical Journal Letters},
}

@article{Amendola:2007rr,
    author = "Amendola, Luca and Kunz, Martin and Sapone, Domenico",
    title = "{Measuring the dark side (with weak lensing)}",
    eprint = "0704.2421",
    archivePrefix = "arXiv",
    primaryClass = "astro-ph",
    doi = "10.1088/1475-7516/2008/04/013",
    journal = "JCAP",
    volume = "04",
    pages = "013",
    year = "2008"
}

@article{PhysRevLett.91.141302,
  title = {Cross-Correlation Tomography: Measuring Dark Energy Evolution with Weak Lensing},
  author = {Jain, Bhuvnesh and Taylor, Andy},
  journal = {Phys. Rev. Lett.},
  volume = {91},
  issue = {14},
  pages = {141302},
  numpages = {4},
  year = {2003},
  month = {Oct},
  publisher = {American Physical Society},
  doi = {10.1103/PhysRevLett.91.141302},
  url = {https://link.aps.org/doi/10.1103/PhysRevLett.91.141302}
}

@article{SpurioMancini:2018apc,
    author = {Spurio Mancini, A. and Reischke, R. and Pettorino, V. and Sch{\"a}fer, B. M. and Zumalac{\'a}rregui, M.},
    title = "{Testing (modified) gravity with 3D and tomographic cosmic shear}",
    eprint = "1801.04251",
    archivePrefix = "arXiv",
    primaryClass = "astro-ph.CO",
    doi = "10.1093/mnras/sty2092",
    journal = "Mon. Not. Roy. Astron. Soc.",
    volume = "480",
    number = "3",
    pages = "3725--3738",
    year = "2018"
}

@article{Hannestad:2006as,
    author = "Hannestad, Steen and Tu, Huitzu and Wong, Yvonne Y. Y.",
    title = "{Measuring neutrino masses and dark energy with weak lensing tomography}",
    eprint = "astro-ph/0603019",
    archivePrefix = "arXiv",
    reportNumber = "MPP-2006-15",
    doi = "10.1088/1475-7516/2006/06/025",
    journal = "JCAP",
    volume = "06",
    pages = "025",
    year = "2006"
}

@ARTICLE{1954ApJ...119..655L,
       author = {{Limber}, D. Nelson},
        title = "{The Analysis of Counts of the Extragalactic Nebulae in Terms of a Fluctuating Density Field. II.}",
      journal = "ApJ",
         year = 1954,
        month = may,
       volume = {119},
        pages = {655},
          doi = {10.1086/145870},
       adsurl = {https://ui.adsabs.harvard.edu/abs/1954ApJ...119..655L}
}

@article{PhysRevLett.13.789,
  title = {Fourth Test of General Relativity},
  author = {Shapiro, Irwin I.},
  journal = {Phys. Rev. Lett.},
  volume = {13},
  issue = {26},
  pages = {789--791},
  numpages = {0},
  year = {1964},
  month = {Dec},
  publisher = {American Physical Society},
  doi = {10.1103/PhysRevLett.13.789},
  url = {https://link.aps.org/doi/10.1103/PhysRevLett.13.789}
}

@article{Hu:2001yq,
    author = "Hu, Wayne and Cooray, Asantha",
    title = "{Gravitational time delay effects on cosmic microwave background anisotropies}",
    eprint = "astro-ph/0008001",
    archivePrefix = "arXiv",
    doi = "10.1103/PhysRevD.63.023504",
    journal = "Phys. Rev. D",
    volume = "63",
    pages = "023504",
    year = "2001"
}

@article{Li:2019qkp,
    author = "Li, Peikai and Dodelson, Scott and Hu, Wayne",
    title = "{Distortions in the Surface of Last Scattering}",
    eprint = "1905.03923",
    archivePrefix = "arXiv",
    primaryClass = "astro-ph.CO",
    doi = "10.1103/PhysRevD.100.043502",
    journal = "Phys. Rev. D",
    volume = "100",
    number = "4",
    pages = "043502",
    year = "2019"
}

@article{Sachs:1967er,
    author = "Sachs, R. K. and Wolfe, A. M.",
    title = "{Perturbations of a cosmological model and angular variations of the microwave background}",
    doi = "10.1007/s10714-007-0448-9",
    journal = "Astrophys. J.",
    volume = "147",
    pages = "73--90",
    year = "1967"
}

@article{Rees:1968zza,
    author = "Rees, M. J. and Sciama, D. W.",
    title = "{Large scale Density Inhomogeneities in the Universe}",
    doi = "10.1038/217511a0",
    journal = "Nature",
    volume = "217",
    pages = "511--516",
    year = "1968"
}

@article{Cai:2010hx,
    author = "Cai, Yan-Chuan and Cole, Shaun and Jenkins, Adrian and Frenk, Carlos S.",
    title = "{Full-sky map of the ISW and Rees-Sciama effect from Gpc simulations}",
    eprint = "1003.0974",
    archivePrefix = "arXiv",
    primaryClass = "astro-ph.CO",
    doi = "10.1111/j.1365-2966.2010.16946.x",
    journal = "Mon. Not. Roy. Astron. Soc.",
    volume = "407",
    pages = "201",
    year = "2010"
}

@article{Cabass:2015xfa,
    author = "Cabass, Giovanni and Gerbino, Martina and Giusarma, Elena and Melchiorri, Alessandro and Pagano, Luca and Salvati, Laura",
    title = "{Constraints on the early and late integrated Sachs-Wolfe effects from the Planck 2015 cosmic microwave background anisotropies in the angular power spectra}",
    eprint = "1507.07586",
    archivePrefix = "arXiv",
    primaryClass = "astro-ph.CO",
    doi = "10.1103/PhysRevD.92.063534",
    journal = "Phys. Rev. D",
    volume = "92",
    number = "6",
    pages = "063534",
    year = "2015"
}

@article{Adamek:2019vko,
    author = "Adamek, Julian and Rasera, Yann and Corasaniti, Pier Stefano and Alimi, Jean-Michel",
    title = "{Ray tracing the integrated Sachs-Wolfe effect through the light cones of the Dark Energy Universe Simulation -- Full Universe Runs}",
    eprint = "1910.03340",
    archivePrefix = "arXiv",
    primaryClass = "astro-ph.CO",
    doi = "10.1103/PhysRevD.101.023512",
    journal = "Phys. Rev. D",
    volume = "101",
    number = "2",
    pages = "023512",
    year = "2020"
}

@article{Beck:2018owr,
    author = "Beck, R{\'o}bert and Csabai, Istv{\'a}n and R{\'a}cz, G{\'a}bor and Szapudi, Istv{\'a}n",
    title = "{The integrated Sachs{\textendash}Wolfe effect in the AvERA cosmology}",
    eprint = "1801.08566",
    archivePrefix = "arXiv",
    primaryClass = "astro-ph.CO",
    doi = "10.1093/mnras/sty1688",
    journal = "Mon. Not. Roy. Astron. Soc.",
    volume = "479",
    number = "3",
    pages = "3582--3591",
    year = "2018"
}

@article{Khosravi:2015boa,
    author = "Khosravi, Shahram and Mollazadeh, Amir and Baghram, Shant",
    title = "{ISW-galaxy cross correlation: a probe of dark energy clustering and distribution of dark matter tracers}",
    eprint = "1510.01720",
    archivePrefix = "arXiv",
    primaryClass = "astro-ph.CO",
    doi = "10.1088/1475-7516/2016/09/003",
    journal = "JCAP",
    volume = "09",
    pages = "003",
    year = "2016"
}

@ARTICLE{1995A&A...301....6C,
       author = {{Cappi}, A.},
        title = "{Gravitational redshift in galaxy clusters.}",
      journal = {Astronomy and Astrophysics},
         year = 1995,
        month = sep,
       volume = {301},
        pages = {6},
       adsurl = {https://ui.adsabs.harvard.edu/abs/1995A&A...301....6C}
}

@article{Wojtak:2011ia,
    author = "Wojtak, Radoslaw and Hansen, Steen H. and Hjorth, Jens",
    title = "{Gravitational redshift of galaxies in clusters as predicted by general relativity}",
    eprint = "1109.6571",
    archivePrefix = "arXiv",
    primaryClass = "astro-ph.CO",
    doi = "10.1038/nature10445",
    journal = "Nature",
    volume = "477",
    pages = "567--569",
    year = "2011"
}

@article{Sadeh:2014rya,
    author = "Sadeh, Iftach and Feng, Low Lerh and Lahav, Ofer",
    title = "{Gravitational Redshift of Galaxies in Clusters from the Sloan Digital Sky Survey and the Baryon Oscillation Spectroscopic Survey}",
    eprint = "1410.5262",
    archivePrefix = "arXiv",
    primaryClass = "astro-ph.CO",
    doi = "10.1103/PhysRevLett.114.071103",
    journal = "Phys. Rev. Lett.",
    volume = "114",
    number = "7",
    pages = "071103",
    year = "2015"
}

@article{Jimeno:2014xma,
    author = "Jimeno, Pablo and Broadhurst, Tom and Coupon, Jean and Umetsu, Keiichi and Lazkoz, Ruth",
    title = "{Comparing gravitational redshifts of SDSS galaxy clusters with the magnified redshift enhancement of background BOSS galaxies}",
    eprint = "1410.6050",
    archivePrefix = "arXiv",
    primaryClass = "astro-ph.CO",
    doi = "10.1093/mnras/stv117",
    journal = "Mon. Not. Roy. Astron. Soc.",
    volume = "448",
    number = "3",
    pages = "1999--2012",
    year = "2015"
}

@article{Zhu:2019cix,
    author = "Zhu, Hongyu and Alam, Shadab and Croft, Rupert A. C. and Ho, Shirley and Giusarma, Elena and Leauthaud, Alexie and Merrifield, Michael",
    title = "{Gravitational redshift profiles of MaNGA BCGs}",
    eprint = "1901.05616",
    archivePrefix = "arXiv",
    primaryClass = "astro-ph.GA",
    month = "1",
    year = "2019"
}

@article{Alonso:2018jzx,
    author = "Alonso, David and Sanchez, Javier and Slosar, An{\v{z}}e",
    collaboration = "LSST Dark Energy Science",
    title = "{A unified pseudo-$C_\ell$ framework}",
    eprint = "1809.09603",
    archivePrefix = "arXiv",
    primaryClass = "astro-ph.CO",
    doi = "10.1093/mnras/stz093",
    journal = "Mon. Not. Roy. Astron. Soc.",
    volume = "484",
    number = "3",
    pages = "4127--4151",
    year = "2019"
}

@article{Alam:2017,
    author = "Alam, Shadab and Zhu, Hongyu and Croft, Rupert A. C. and Ho, Shirley and Giusarma, Elena and Schneider, Donald P.",
    title = "{Relativistic distortions in the large-scale clustering of SDSS-III BOSS CMASS galaxies}",
    eprint = "1709.07855",
    archivePrefix = "arXiv",
    primaryClass = "astro-ph.CO",
    doi = "10.1093/mnras/stx1421",
    journal = "Mon. Not. Roy. Astron. Soc.",
    volume = "470",
    number = "3",
    pages = "2822--2833",
    year = "2017"
}

@article{Saga:2021,
    author = "Saga, Shohei and Taruya, Atsushi and Breton, Michel-Andr{\`e}s and Rasera, Yann",
    title = "{Detectability of the gravitational redshift effect from the asymmetric galaxy clustering}",
    eprint = "2109.06012",
    archivePrefix = "arXiv",
    primaryClass = "astro-ph.CO",
    reportNumber = "YITP-21-93",
    doi = "10.1093/mnras/stac186",
    journal = "Mon. Not. Roy. Astron. Soc.",
    volume = "511",
    number = "2",
    pages = "2732--2754",
    year = "2022"
}

@article{SobralBlanco:2022,
    author = "Sobral-Blanco, Daniel and Bonvin, Camille",
    title = "{Measuring the distortion of time with relativistic effects in large-scale structure}",
    eprint = "2205.02567",
    archivePrefix = "arXiv",
    primaryClass = "astro-ph.CO",
    doi = "10.1093/mnrasl/slac124",
    journal = "Mon. Not. Roy. Astron. Soc.",
    volume = "519",
    number = "1",
    pages = "L39--L44",
    year = "2023"
}

@article{Lepori:2024,
    author = "Lepori, F. and Schulz, S. and Tutusaus, I. and Breton, M.-A. and Saga, S. and Viglione, C. and Adamek, J. and Bonvin, C. and Dam, L. and Fosalba, P. and others",
    collaboration = "Euclid",
    title = "{Euclid: Relativistic effects in the dipole of the 2-point correlation function}",
    eprint = "2410.06268",
    archivePrefix = "arXiv",
    primaryClass = "astro-ph.CO",
    year = "2024"
}

@article{Dam:2025,
    author = "Dam, Lawrence and Bonvin, Camille",
    title = "{Gravitational redshift from large-scale structure: nonlinearities, anti-symmetries, and the dipole}",
    eprint = "2506.22431",
    archivePrefix = "arXiv",
    primaryClass = "astro-ph.CO",
    year = "2025"
}

@article{Adamek:2026vof,
    author = "Adamek, Julian and Christiansen, {\O}yvind",
    title = "{gevolution 2.0: GPU-accelerated relativistic N-body simulations for cosmology}",
    eprint = "2607.17929",
    archivePrefix = "arXiv",
    primaryClass = "astro-ph.IM",
    month = "7",
    year = "2026"
}
\end{document}